\documentclass{SCIS2026}

\begin{document}

\ArticleType{RESEARCH PAPER}
\Year{2025}
\Month{January}
\Vol{68}
\No{1}
\DOI{}
\ArtNo{}
\ReceiveDate{}
\ReviseDate{}
\AcceptDate{}
\OnlineDate{}
\AuthorMark{}
\AuthorCitation{}

\title{Cross-reality location privacy protection in 6G-enabled vehicular metaverses: an LLM-enhanced hybrid generative diffusion model-based approach}{Luo X F, He J Y, Kang J W, et al. Cross-reality location privacy protection in 6G-enabled vehicular metaverses: an LLM-enhanced hybrid generative diffusion model-based approach}

\author[1]{Xiaofeng Luo}{}
\author[1]{Jiayi He}{}
\author[1,2]{Jiawen Kang}{{kavinkang@gdut.edu.cn}}
\author[3]{Ruichen Zhang}{}
\author[1]{Zhaoshui He}{}
\author[4]{\\Ekram Hossain}{}
\author[5]{Dong In Kim}{}

\address[1]{School of Automation, Guangdong University of Technology, Guangzhou {\rm 510006}, China}
\address[2]{State Key Laboratory of Integrated Services Networks, Xidian University, Xi’an {\rm 710071}, China}
\address[3]{College of Computing and Data Science, Nanyang Technological University, Singapore {\rm 639798}, Singapore}
\address[4]{Department of Electrical and Computer Engineering, University of Manitoba, Winnipeg, {\rm MB R3T 2N2}, Canada}
\address[5]{Department of Electrical and Computer Engineering, Sungkyunkwan University, Suwon 16419, South Korea}

\abstract{The emergence of 6G-enabled vehicular metaverses enables Autonomous Vehicles (AVs) to operate across physical and virtual spaces through space-air-ground-sea integrated networks. The AVs can deploy Artificial Intelligence (AI) agents powered by large AI models on edge servers as personalized assistants to support intelligent driving decision-making and enhanced on-board experiences. However, such cross-reality interactions between physical vehicle states and virtual AI agent deployments may cause serious location privacy risks, as adversaries can infer AV trajectories by correlating the locations reported when AVs request Location-Based Services (LBS) in reality with the locations of the edge servers on which their corresponding AI agents are deployed in virtuality. To address this challenge, we design a cross-reality location privacy protection framework based on hybrid actions, including continuous location perturbation in reality and discrete privacy-aware AI agent migration in virtuality. In this framework, a new privacy metric, termed cross-reality location entropy, is proposed to effectively quantify the privacy levels of AVs. Based on this metric, we formulate an optimization problem to optimize the hybrid actions, focusing on achieving a balance between location privacy protection, service latency reduction, and quality of service maintenance. To solve the complex mixed-integer problem, we develop a novel \underline{\textbf{L}}LM-enhanced \underline{\textbf{H}}ybrid \underline{\textbf{D}}iffusion \underline{\textbf{P}}roximal \underline{\textbf{P}}olicy \underline{\textbf{O}}ptimization (LHDPPO) algorithm, which integrates Large Language Model (LLM)-driven informative reward design to enhance environment understanding with dual-chain Generative Diffusion Model (GDM)-based policy exploration to handle high-dimensional action spaces, thereby enabling reliable determination of optimal hybrid actions. Extensive experiments on real-world datasets demonstrate that the proposed framework effectively mitigates cross-reality location privacy leakage for AVs while maintaining strong user immersion within 6G-enabled vehicular metaverse scenarios.}

\keywords{Location privacy, vehicular metaverse, large AI model, generative diffusion model, 6G networks, agentic AI}

\maketitle

\section{Introduction}
The vehicular metaverse is a mobility paradigm that blends physical and virtual spaces to enhance intelligence and interactivity in vehicular networks~\cite{xie2025towards,chen2025efficient}. With the advancement of Sixth-Generation (6G) cellular technology, 6G-enabled vehicular metaverses have emerged to provide ultra-reliable and low-latency connectivity across space-air-ground-sea integrated networks, enabling uninterrupted and high-quality services for on-board users~\cite{kang2025diffusion,cui2025overview}.  In this paradigm, due to the limited computing and storage resources of Autonomous Vehicles (AVs), AVs typically deploy Artificial Intelligence (AI) agents powered by large AI models as their personalized assistants on edge servers, which can be terrestrial Roadside Units (RSUs) or Low-Earth Orbit (LEO) satellites~\cite{wang2025uplink,kang2024blockchain}. By hosting AI agents and executing computation-intensive tasks, these edge servers enable resource-constrained AVs to access large AI model-based services, thereby enhancing driving safety and on-board experience~\cite{xue2025large}. In this way, AVs exhibit agentic AI characteristics by leveraging perception, reasoning, execution, and continuous learning capabilities to interact with physical and virtual spaces in an iterative and context-aware manner~\cite{zhang2026toward}.

Although the 6G-enabled vehicular metaverse empowered by agentic AI enables immersive services and intelligent decision-making across physical and virtual spaces, privacy risks become more prominent in such tightly coupled cross-reality systems, i.e., systems linking physical vehicular states with virtual metaverse infrastructures, than in traditional vehicular networks~\cite{auda2023scoping,hiroi2025cross}. Specifically, the immersion of on-board users within the vehicular metaverse is highly correlated with two key factors, i.e., reported location precision and service response latency~\cite{luo2025incentivizing}. On the one hand, agentic AI-driven AVs in the physical space frequently request Location-Based Services (LBS), such as Augmented Reality (AR) navigation and AR interactive games, where higher reported location precision ensures better Quality of Service (QoS) but also incurs a higher risk of location privacy leakage~\cite{min2023personalized}. On the other hand, due to the continuous mobility of AVs, their corresponding AI agents in the virtual space are frequently migrated among edge servers in alignment with physical vehicle movements to maintain physical–virtual synchronization and low-latency service provisioning, which further amplifies privacy risks~\cite{chen2025efficient,wang2023location}. Consequently, collusive adversaries can exploit their prior knowledge of AVs' routine mobility patterns, such as regular commuting routes, to launch inference attacks that jointly reconstruct vehicle trajectories and AI agent migration patterns, thereby accurately tracking target AVs~\cite{luo2023privacy}. This cross-reality linkage attack is more severe and harder to mitigate than traditional single-domain privacy attacks, posing substantial threats to the location privacy of on-board users.

Existing research on cross-reality location privacy preservation in vehicular metaverses has primarily focused on anonymity-based schemes~\cite{kang2024blockchain,luo2025incentivizing}, which aim to obscure the linkage between vehicle identities and locations. However, these approaches often suffer from significant performance degradation in low-traffic environments and fail to meet the stringent privacy requirements of AVs~\cite{ullah2025location}. These limitations highlight the need for a general solution capable of protecting cross-reality location privacy in 6G-enabled vehicular metaverses. Moreover, most existing location privacy metrics, such as anonymity set size, adversary success rate, geo-indistinguishability-based entropy, and migration-based location entropy, consider only a single dimension, typically confined to the physical space or a narrow service-migration process~\cite{emara2017safety,andres2013geo,li2022quantifying,wang2023location}. Therefore, they cannot accurately capture privacy leakage risks arising from cross-reality interactions. Consequently, a more comprehensive metric is required to effectively evaluate location privacy under cross-reality inference attacks. Furthermore, strengthening privacy protection at a large scale inevitably increases service response latency and degrades QoS, thereby significantly diminishing user immersion in metaverse environments. As a result, achieving an effective balance among cross-reality location privacy protection, latency reduction, and QoS maintenance remains a challenging and open research problem.

To address the above challenges, this paper proposes a user-centric cross-reality location privacy protection framework based on hybrid action decisions. The main contributions of this paper are summarized as follows.
\begin{itemize}
    \item We present a cross-reality location privacy protection \textbf{framework} for 6G-enabled vehicular metaverses to mitigate the risk of privacy leakage caused by spatiotemporal correlations between physical and virtual spaces. The framework integrates a hybrid action \textbf{mechanism} that combines continuous location perturbation in reality with discrete privacy-aware AI agent migration in virtuality to preserve the location privacy of agentic AI-driven AVs.
    \item In the proposed framework, a new privacy \textbf{metric} named cross-reality location entropy is designed to quantify the real-time location privacy levels of AVs under cross-reality inference attacks. The metric is derived from a multi-condition Bayesian posterior distribution and captures the uncertainty of vehicle locations introduced by hybrid action strategies. Based on this metric, a mixed-integer optimization problem that jointly considers cross-reality location privacy, service response latency, and QoS loss is formulated to optimize the hybrid action mechanism.
    \item To solve the formulated complex problem in vehicular metaverse environments, a novel \underline{\textbf{L}}LM-enhanced \underline{\textbf{H}}ybrid \underline{\textbf{D}}iffusion \underline{\textbf{P}}roximal \underline{\textbf{P}}olicy \underline{\textbf{O}}ptimization (LHDPPO) \textbf{algorithm} is proposed, which incorporates Large Language Model (LLM)-driven informative reward design for enhanced environment understanding and Generative Diffusion Model (GDM)-based policy exploration for improved learning performance in high-dimensional action spaces.
    \item Extensive experiments using real-world taxi mobility and base station datasets demonstrate that the proposed framework achieves superior performance in terms of cross-reality location privacy preservation, service response latency reduction for immersive services, and QoS maintenance when compared with existing baseline schemes.
\end{itemize}

The remainder of this paper is organized as follows. Section~\ref{Related Work} reviews related studies on cross-reality location privacy protection, privacy metrics, and GDM-based reinforcement learning. Section~\ref{Scheme} presents the proposed cross-reality location privacy protection framework. Section~\ref{System model} presents the system model and formulates the hybrid action optimization problem, while Section~\ref{DRL_diffusion} details the proposed LHDPPO algorithm. Section~\ref{Performance Evaluation} reports the experimental results, followed by the conclusion in Section~\ref{Conclusion}.

\section{Related work}
\label{Related Work}
\subsection{Location privacy protection in cross-reality systems}
Location privacy protection has become an active research topic in cross-reality systems, where existing solutions are generally classified into encryption-based, anonymization-based, and obfuscation-based schemes~\cite{liu2018location}. For encryption-based approaches, Liu et al.~\cite{liu2023blocksc} proposed BlockSC, a blockchain-empowered spatial crowdsourcing framework that preserves location privacy in metaverses through spatial data encryption. Although effective in securing sensitive information, the heavy computational overhead of cryptographic operations limits its suitability for real-time services in 6G-enabled vehicular metaverses. Anonymization-based schemes aim to weaken the association between users and their trajectories by concealing real identities. For instance, Luo et al.~\cite{luo2023privacy} developed a dual pseudonym change scheme to achieve identity anonymization in vehicular metaverses, but its effectiveness degrades in sparse traffic areas, and it cannot fully realize location concealment. Obfuscation-based schemes protect privacy by perturbing reported positions to reduce adversarial inference accuracy. For instance, Wang et al.~\cite{wang2025sustainable} proposed a perturbed sliding task queue algorithm based on differential privacy to protect sensitive user location information prior to task offloading in a digital twin architecture, yet it considers only physical space location perturbation and neglects interactions in the virtual space. Therefore, there is still a lack of practical solutions that are capable of preserving cross-reality location privacy in 6G-enabled vehicular metaverses.

\subsection{Location privacy metrics}
Location privacy metrics provide quantitative tools for evaluating protection schemes in mobile networks. Emara et al.~\cite{emara2017safety} showed that widely used metrics in VANETs, such as anonymity set size, entropy, and adversary success rate, each capture only a partial aspect of privacy and may yield inconsistent evaluations across scenarios. Li et al.~\cite{li2022quantifying} proposed distance and visibility-based metrics for navigation services, but the metrics remain confined to physical trajectories and ignore virtual space privacy leakage. To model inference from service migration, an entropy-based metric proposed in~\cite{wang2023location} quantifies location uncertainty under MEC service placement, although it considers only location exposure during service migration and neglects sensitive broadcast information in the physical space. More recently, Kang et al.~\cite{kang2024blockchain} introduced the Degree of Privacy Entropy (DoPE) to assess pseudonym changes in vehicular metaverses, yet DoPE targets identity anonymity within pseudonym-based schemes and does not apply to perturbation-based mechanisms. Overall, existing metrics evaluate privacy within a single domain or narrow application scope and cannot capture cross-reality linkability between broadcast messages and metaverse activities, revealing the need for location privacy metrics tailored to 6G-enabled vehicular metaverses.

\subsection{GDM-based reinforcement learning for decision optimization}
Recent progress in GDMs has demonstrated strong potential in solving complex optimization problems. Du et al.~\cite{du2024enhancing} surveyed GDM-based Deep Reinforcement Learning (DRL) and presented a power allocation example showing that GDMs can capture high-dimensional structures and generate high-quality decisions. Building on this insight, Ning et al.~\cite{ning2025diffusion} proposed a two-level GDM-based soft actor critic framework for effective resource management, but their method supports only a single action type and has limited applicability. Hence, Kang et al.~\cite{kang2025hybrid} introduced a hybrid GDM-based DRL algorithm for digital twin migration in vehicular metaverses, jointly generating migration and offloading actions. However, their method relies on traditional RL relaxation techniques to handle hybrid actions, which may destabilize learning and cause suboptimal convergence during training. To further improve hybrid action modeling, Wang et al.~\cite{wang2025uplink} embedded GDMs into a hybrid proximal policy optimization algorithm through a parameterized action formulation to optimize spectral efficiency in LEO satellite networks. Although effective for parameterized decision tasks, this structure imposes mutual exclusivity among actions and limits exploration of new hybrid action pairs. Moreover, existing GDM-based DRL algorithms for optimization still depend entirely on manually crafted reward functions, which may neglect the potentially important components in intricate cross-reality environments. These limitations highlight the need for a new GDM-based DRL algorithm that decouples hybrid actions and leverages large AI models to design informative reward functions automatically.

\begin{figure*}[t]
    \centerline{\includegraphics[width=0.85\textwidth]{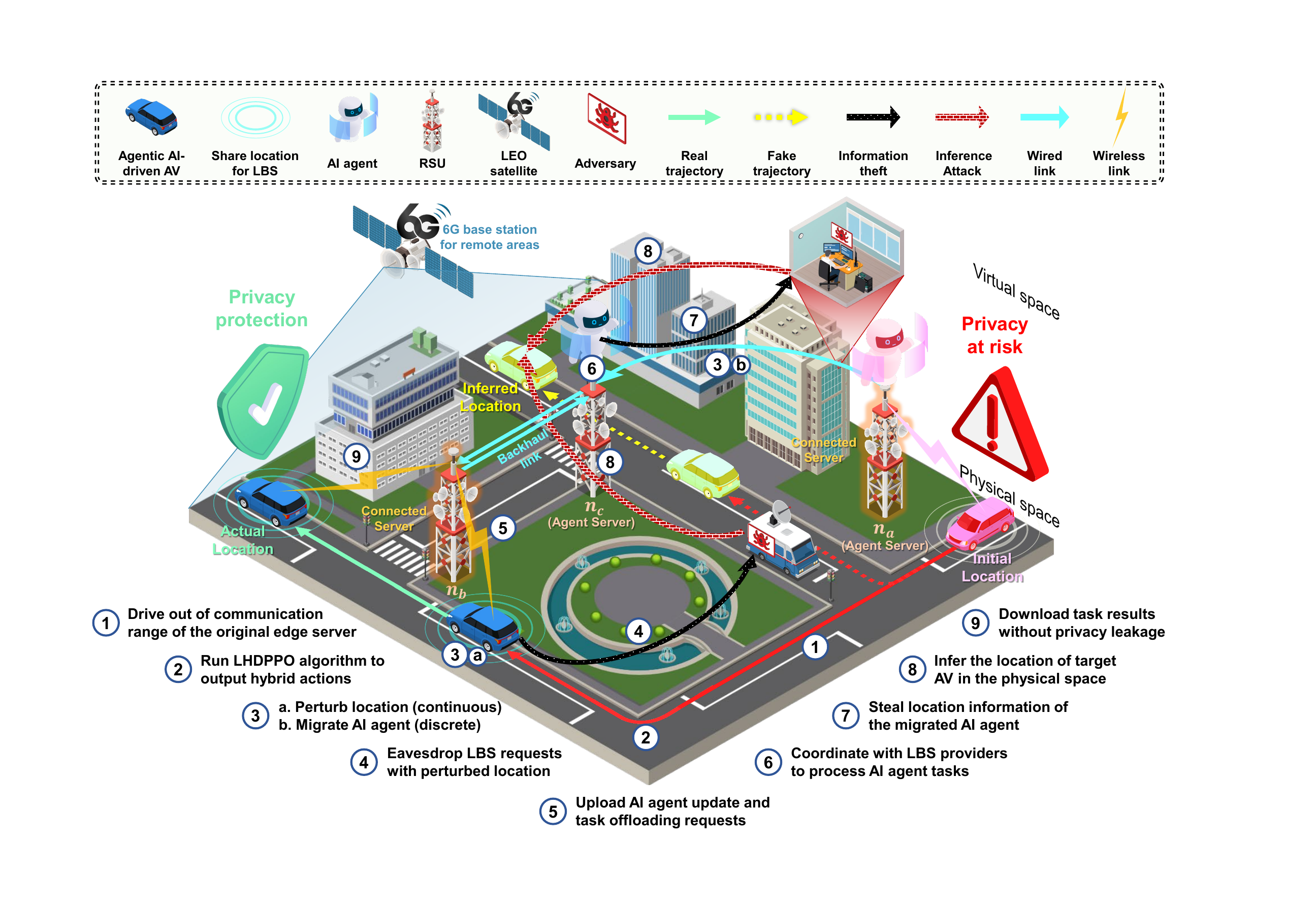}}
    \caption{(Color online) Workflow of the proposed cross-reality location privacy protection framework based on hybrid actions in 6G-enabled vehicular metaverses. The sequentially numbered arrows depict the main processes: a) The target AV at privacy risk moves and executes hybrid actions, i.e., continuous location perturbation and discrete privacy-aware AI agent migration (Steps \textcircled{1} to \textcircled{3}). b) The AV requests LBS and offloads AI agent tasks to edge servers, while malicious adversaries exploit illegally obtained target-AV information to launch inference attacks (Steps \textcircled{4} to \textcircled{8}). c) The on-board user obtains immersive metaverse applications without compromising location privacy (Step \textcircled{9}).}
    \label{fig1}
\end{figure*}

\section{Cross-reality location privacy protection framework in 6G-enabled vehicular metaverses}
\label{Scheme}
In this section, we propose a cross-reality location privacy framework based on hybrid action mechanisms in 6G-enabled vehicular metaverses. As illustrated in Figure~\ref{fig1}, each AV in reality is associated with an AI agent in virtuality, which is hosted on one of the edge servers. First, the key components of the 6G-enabled vehicular metaverse are described in the network model. Subsequently, the core hybrid action mechanism is presented, focusing on safeguarding location privacy across both physical and virtual spaces.

\subsection{Network model}
\label{network_model}
\begin{itemize}
    \item \textbf{Agentic AI-driven AVs:} Agentic AI-driven AVs operating in the physical space act as primary participants in 6G-enabled vehicular metaverses. Each AV can request customized LBS, such as AR navigation, to enhance the on-board experience throughout the journey. Although these services are essential for user immersion, the associated service requests inevitably disclose sensitive location information, which can be exploited by adversaries to track AV movements~\cite{min2023personalized}.

    \item \textbf{AI agents:} AI agents residing in the virtual space act as intelligent assistants for their corresponding AVs. By directly interacting with other AI agents and LBS providers, AI agents empowered by large AI models can assist AVs in processing computation-intensive tasks and providing decision-support functionalities. However, frequent interactions and continuous migrations of AI agents may inadvertently leak location-related information~\cite{luo2023privacy}. To tackle these privacy threats, the large AI model-based AI agents can support AVs with agentic AI capabilities to determine hybrid actions for cross-reality location privacy protection~\cite{yang2025privacy}.

    \item \textbf{Edge Servers:} Edge servers function as the bridge between the physical and virtual spaces in 6G-enabled vehicular metaverses. Equipped with larger computing and storage resources than AVs, they manage the full lifecycle of AI agents, including creation, synchronization, execution, and migration. In the 6G-enabled vehicular metaverse, edge servers can be terrestrial RSUs or LEO satellites, where satellite nodes act as 6G base stations to compensate for poor terrestrial connectivity in remote areas~\cite{sheng2023coverage}.
    
    \item \textbf{Adversaries:} Adversaries may exist in both physical and virtual spaces and are considered to possess prior knowledge about target AVs. This consideration is reasonable, as the driving routes and social interaction patterns of an AV often exhibit temporal regularity~\cite{wang2023location}. In the physical space, adversaries can eavesdrop on sensitive information broadcast by AVs when they request LBS. In the virtual space, adversaries may be malicious AI agents or metaverse service providers aiming to infer user locations, particularly during AI agent migrations. Furthermore, adversaries across physical and virtual spaces may share information to launch collusive inference attacks to associate an AV with its AI agent, posing a significant threat to the location privacy of on-board users~\cite{luo2023privacy}.
\end{itemize}

\subsection{Hybrid action-based cross-reality privacy protection mechanism}
To conceal user location in 6G-enabled vehicular metaverses, the proposed cross-reality location privacy protection framework adopts a hybrid action mechanism consisting of a continuous action for location perturbation in the physical space and a discrete action for privacy-aware AI agent migration in the virtual space. See the details below.

\subsubsection{Continuous action---location perturbation in reality}
We consider an observation area containing multiple agentic AI-driven AVs, multiple RSUs, and one LEO satellite. Specifically, we denote an AV set $\mathbb{M}= \{1,\ldots,m,\ldots,M\}$ with a total of $M$ AVs and an RSU set $\mathbb{N}= \{1,\ldots,n,\ldots,N\}$ with a total of $N$ RSUs, while the index of the LEO satellite is denoted as $S$. In the physical space, location perturbation (also called location obfuscation) can be applied to AV trajectories to reduce the risk of direct location exposure~\cite{ullah2025location}. Each AV introduces controlled perturbations to its reported coordinates when requesting LBS. Notably, larger perturbations increase privacy by enlarging the uncertainty of an adversary’s inference but also weaken QoS~\cite{min2023personalized}. Hence, the perturbation intensity can be modeled as a continuous action variable that is adjusted based on the privacy and QoS requirements of AVs.

The planar Laplacian perturbation mechanism proposed in~\cite{andres2013geo} provides a principled approach to achieving geo-indistinguishability based on differential privacy. This method samples random noise from a two-dimensional Laplace distribution and perturbs the original location to satisfy $\epsilon$-geo-indistinguishability, where $\epsilon$ denotes the privacy budget. Inspired by this approach, the proposed cross-reality location privacy protection framework adopts a similar mechanism to implement continuous location perturbation in the physical space.

Let $l_m^t=(x_m^t,y_m^t)$ denote the true location of AV $m$ at time slot $t$, where $x_m^t$ and $y_m^t$ represent the latitude and longitude, respectively. In a polar coordinate system centered at $l_m^t$, any relative location $l_x$ can be represented as $(r_m^t,\theta_m^t)$, where $r_m^t$ denotes the perturbation radius and $\theta_m^t$ denotes the perturbation angle with respect to the horizontal axis. According to~\cite{andres2013geo}, the Probability Density Function (PDF) of the polar Laplacian distribution is
\begin{equation}
\label{laplace}
    D_{\epsilon}(r_m^t,\theta_m^t)=\frac{\epsilon^2}{2\pi}r_m^t e^{-\epsilon r_m^t}.
\end{equation}
Eq. (\ref{laplace}) reflects that the perturbation is jointly determined by the continuous random variables $r_m^t$ and $\theta_m^t$. Hence, to perform location perturbation at time slot $t$, AV $m$ only needs to determine a continuous-action policy $\boldsymbol{c}_m^t=\{r_m^t,\theta_m^t\}$. Given the selected actions $r_m^t$ and $\theta_m^t\in[0,2\pi)$, the perturbed location $\tilde{l}_m^t$ is computed as
\begin{equation}
\label{perturbed_location}
    \tilde{l}_m^t=(\tilde{x}_m^t,\tilde{y}_m^t)=(x_m^t+r_m^t\cos{\theta_m^t},y_m^t+r_m^t\sin{\theta_m^t}).
\end{equation}
The detailed procedure for determining $\boldsymbol{c}_m^t$ with the LLM-enhanced learning approach is presented in Section~\ref{DRL_diffusion}.

\subsubsection{Discrete action---privacy-aware AI agent migration in virtuality}
\label{discrete_action}
In the virtual space, AI agent migration not only sustains continuous service provisioning under AV mobility but also serves as an effective mechanism for enhancing location privacy~\cite{wang2023location}. By dynamically changing the hosting edge server of an AI agent, the one-to-one correspondence between an AV and a specific server can be disrupted, thereby introducing uncertainty into the inferred locations of AVs and reducing privacy risks. However, if an AV always migrates its AI agent to the server offering the strongest signal strength, an adversary continuously monitoring the target AI agent can still reconstruct the real trajectory of the target AV. Conversely, migrating the AI agent to a distant edge server may increase service response latency and degrade user immersion in 6G-enabled vehicular metaverses. To address this tradeoff, privacy-aware AI agent migration is modeled as a discrete action, where an AV optimally selects whether and where to migrate its agent among edge server candidates, achieving a dynamic balance between location privacy protection and service latency reduction.

Let $c_m^t$ denote the index of the edge server connected to AV $m$ at time slot $t$. In the 6G-enabled vehicular metaverse, each AV is connected to exactly one edge server in each time slot, which is expressed as
$\mathbb{I}(c_m^t\in\mathbb{N})+\mathbb{I}(c_m^t=S)=1$,
where $\mathbb{I}(\cdot)$ is the indicator function. Meanwhile, let $e_m^t$ denote the index of the edge server that hosts the AI agent of AV $m$, referred to as the agent server. As illustrated in Figure~\ref{fig1}, the glowing RSU $n_a\in\mathbb{N}$ represents the edge server providing the strongest signal to AV $m$ at its initial position, serving both as the connected edge server and the initial agent server. When AV $m$ moves out of the coverage range of RSU $n_a$, it connects to another RSU $n_b\in\mathbb{N}$ that offers the strongest signal at the new position. However, the connected server $n_b$ is not necessarily selected as the agent server. After evaluating the current location privacy level, AV $m$ can instead choose an alternative RSU $n_c$ ($n_c\in\mathbb{N}$) to balance privacy protection and service latency based on its discrete action decision policy $\boldsymbol{d}_m^t=n_c$.

Therefore, to protect location privacy in the virtual space, the agentic AI-driven AV $m$ can determine its discrete action policy $\boldsymbol{d}_m^t=\{e_m^t \mid \mathbb{I}(e_m^t\in\mathbb{N})+\mathbb{I}(e_m^t=S)=1\}$ for privacy-aware AI agent migration in time slot $t$. The detailed procedure for determining $\boldsymbol{d}_m^t$ with the LLM-enhanced learning approach is presented in Section~\ref{DRL_diffusion}.

\section{System model}
\label{System model}
Based on the above analysis, the core of the proposed framework for achieving cross-reality location privacy protection is to determine the optimal hybrid action, striking a balance between location privacy protection and user immersion. To achieve this, we design a new privacy metric named cross-reality location entropy to evaluate privacy levels under the hybrid action mechanisms. Meanwhile, the service response latency and QoS loss are explicitly modeled. By integrating these components, the overall optimization problem is formulated in Section \ref{optimization_problem}.

\subsection{New privacy metric: cross-reality location entropy}
Before determining the hybrid action for cross-reality location privacy protection, it is essential for AV $m$ to evaluate its current privacy level. When the assessed privacy level is high, AV $m$ can prioritize its immersive experience within the metaverse, meaning that it can request LBS with accurate locations and deploy its AI agent on a nearby edge server. In contrast, AV $m$ can adopt a more aggressive protection strategy, requesting LBS with largely perturbed positions and migrating its AI agent to a more distant edge server.

As discussed in Section \ref{network_model}, adversaries can obtain partial historical movement trajectories of target AVs. With this side information, an adversary can analyze the migration pattern of the target AV and conduct a Bayesian location inference attack, which is a type of knowledge-based attack~\cite{wang2023location}. Combining past movement traces and migration patterns, adversaries can build a mapping relationship between an AV's location in reality and its AI agent's location in virtuality, learning the probability distribution of the AV's real-time position. Specifically, let $\Pr(\breve{l}_m^t=p_a)$, $\Pr(\tilde{l}_m^t=p_b)$, and $\Pr(\hat{l}_m^t=p_c)$ denote the probabilities that the inferred AV location, perturbed AV location, and corresponding AI agent location of AV $m$ are at positions $p_a$, $p_b$, and $p_c$, respectively. Then, the posterior distribution of AV $m$’s location is obtained by observing the perturbed AV location in reality and the AI agent location in virtuality based on the multi-condition Bayes' theorem, given by
\begin{equation}
\label{posterior_distribution}
    \Pr(\breve{l}_m^t=p_a\mid \tilde{l}_m^t=p_b,\hat{l}_m^t=p_c)=\frac{\Pr(\breve{l}_m^t=p_a,\tilde{l}_m^t=p_b,\hat{l}_m^t=p_c)}{\Pr(\tilde{l}_m^t=p_b,\hat{l}_m^t=p_c)}=\frac{\Pr(\tilde{l}_m^t=p_b,\hat{l}_m^t=p_c\mid \breve{l}_m^t=p_a) \Pr(\breve{l}_m^t=p_a)}{\sum_{p_x\in\mathcal{P}_m^t} \Pr(\tilde{l}_m^t=p_b,\hat{l}_m^t=p_c\mid l_m^t=p_x) \Pr(l_m^t=p_x)},
\end{equation}
where $\mathcal{P}_m^t$ is the set of potential physical locations of AV $m$ in time slot $t$. Since the perturbed location and AI agent location are each determined solely by the hybrid action strategy of AV $m$, the two conditional probabilities can be treated as mutually independent when the physical AV location is given. In this case, the posterior distribution in Eq. (\ref{posterior_distribution}) is transformed into
\begin{equation}
\Pr(\breve{l}_m^t=p_a\mid\tilde{l}_m^t=p_b,\hat{l}_m^t=p_c)
=\frac{\Pr(\tilde{l}_m^t=p_b\mid \breve{l}_m^t=p_a)\Pr(\hat{l}_m^t=p_c\mid \breve{l}_m^t=p_a)\Pr(\breve{l}_m^t=p_a)}
{\sum_{p_x\in\mathcal{P}_m^t}\Pr(\tilde{l}_m^t=p_b\mid l_m^t=p_x)\Pr(\hat{l}_m^t=p_c\mid l_m^t=p_x)\Pr(l_m^t=p_x)}.
\end{equation}
To assess the privacy levels of AVs in 6G-enabled vehicular metaverses, we design a new physical-virtual location privacy metric based on information entropy, referred to as cross-reality location entropy.
\begin{definition}
    \textup{\textbf{(Cross-reality Location Entropy)}} \textit{In the 6G-enabled vehicular metaverse, the cross-reality location entropy of an AV $m$ at time slot $t$ is defined as} 
\end{definition}
\begin{equation}
\label{location_entropy}
    E_m^t=-\sum_{p_a\in\mathcal{P}_m^t}\Pr(\breve{l}_m^t=p_a\mid \tilde{l}_m^t=p_b,\hat{l}_m^t=p_c) \log_2 \bigg(\Pr(\breve{l}_m^t=p_a\mid \tilde{l}_m^t=p_b,\hat{l}_m^t=p_c)\bigg).
\end{equation}

Notably, the cross-reality location entropy in Eq. (\ref{location_entropy}) is established based on the concept of entropy in information theory~\cite{renyi1961measures}. A larger entropy value implies a lower accuracy of the adversary's inference regarding the actual location of the target AV, indicating a lower risk of location privacy leakage. With this metric, AVs can evaluate their location privacy levels across physical and virtual spaces, thereby guiding hybrid action mechanisms to maintain an appropriate balance between location privacy and user immersion in 6G-enabled vehicular metaverses.

\subsection{Service response latency model}
\subsubsection{Communication model}
The communication model describes the wireless transmission processes between AVs and edge servers. To maintain accurate physical–virtual synchronization and receive high-quality metaverse services, AVs should upload real-time sensing data to edge servers for updating their large AI model-empowered AI agents and offloading computation tasks through uplink channels~\cite{luo2025incentivizing}. Consider the case where the AI agent of AV $m$ is deployed on RSU $n$ in the virtual space during time slot $t$, i.e., $e_m^t=n$. Let $\hat{l}_m^t=l_n=\{x_n,y_n\}$ denote the position of agent server $e_m^t$, which corresponds to the fixed geographical coordinates of RSU $n$, with $x_n$ and $y_n$ representing its latitude and longitude, respectively. The latitudinal difference between AV $m$ and RSU $n$ is thus given by $\Delta x_{m,n}^t=\frac{\pi}{180}\left(x_m^t-x_n\right)$, while the longitudinal difference is $\Delta y_{m,n}^t=\frac{\pi}{180}(y_m^t-y_n)$. Hence, the haversine of the central angle $\theta_{m,n}^t$ (in radians) is
\begin{equation}
\label{have}
    hav(\theta_{m,n}^t)=\sin^2(\frac{\theta_{m,n}^t}{2})=\sin^2 (\frac{\Delta x_{m,n}^t}{2}) +\cos(\frac{\pi}{180}x_m^t)\cos(\frac{\pi}{180} x_n) \sin^2 (\frac{\Delta y_{m,n}^t}{2}).
\end{equation}
From Eq.~(\ref{have}), we know that $\theta_{m,n}^t = 2\arcsin{\sqrt{hav(\theta_{m,n}^t)}}$. Then, the actual distance between AV $m$ and RSU $n$ is
\begin{equation}
\label{distance_formula}
    d_{m,n}^t=R\theta_{m,n}^t=2R\arcsin{\sqrt{hav(\theta_{m,n}^t)}},
\end{equation}
where $R=6371$ km denotes the radius of the Earth.

For vehicular systems, the Rayleigh fading model is commonly used to characterize the wireless channel between RSUs and AVs~\cite{chen2025efficient,wu2025energy}. The channel gain is defined as $h_{m,n}^t=A\left[\frac{c}{4\pi f_c d_{m,n}^t}\right]^\xi$, where $c=3\times10^8$ m/s is the speed of light, $A$ is the antenna gain, $f_c$ is the carrier frequency, and $\xi$ is the path loss factor~\cite{chen2025efficient}. Therefore, the achievable uplink data rate from AV $m$ to the server $c_m^t$ is
\begin{equation}
R_{m,n}^{\text{up}}(t)=B^{\text{up}}_{m,n}\log_2\left(1+\frac{p_m h_{m,n}^t}{N_0}\right),
\end{equation}
where $B^{\text{up}}_{m,n}$ is the uplink bandwidth allocated to AV $m$ when connected to RSU $n$, $p_m$ is the transmit power of AV $m$, and $N_0$ is the average noise power. Considering two types of edge servers, the uplink communication latency is
\begin{equation}
    L_{m}^{\text{up}}(t)=\mathbb{I}(c_m^t\in\mathbb{N})\frac{D_{m}^{\text{up}}(t)}{R_{m,n}^{\text{up}}(t)}+\mathbb{I}(c_m^t=S)\frac{D_{m}^{\text{up}}(t)}{R_{\text{sat}}^{\text{up}}},
\end{equation}
where $D_{m}^{\text{up}}(t)$ is the amount of data uploaded by AV $m$ for AI agent updates and task offloading, and $R_{\text{sat}}^{\text{up}}$ denotes the average ground-to-satellite upload speed in Starlink\footnote{\url{https://www.starlink.com/technology}}.

In addition to uplink transmission, AVs also receive customized AI agent task results from the RSU via the downlink channel to obtain personalized metaverse services (e.g., AR navigation to a preferred destination). Considering channel reciprocity, the uplink and downlink share identical path-loss characteristics~\cite{kang2025hybrid}. Accordingly, the downlink data rate is
$R_{m,n}^{\text{down}}(t)=B^{\text{down}}_{m,n}\log_2\left(1+\frac{p_n h_{m,n}^t}{N_0}\right)$,
where $B^{\text{down}}_{m,n}$ denotes the downlink bandwidth allocated to AV $m$ when connected to edge server $c_m^t$, and $p_n$ is the transmit power of RSU $n$~\cite{zhang2025personalizing}. To receive AI agent task results of size $D_{m}^{\text{down}}(t)$, the downlink communication latency in time slot $t$ is given by
\begin{equation}
    L_{m}^{\text{down}}(t)=\mathbb{I}(c_m^t\in\mathbb{N})\frac{D_{m}^{\text{down}}(t)}{R_{m,n}^{\text{down}}(t)}+\mathbb{I}(c_m^t=S)\frac{D_{m}^{\text{down}}(t)}{R_{\text{sat}}^{\text{down}}},
\end{equation}
where $R_{\text{sat}}^{\text{down}}$ denotes the average ground-to-satellite download speed in Starlink\footnotemark[1].

Since the connected edge server may differ from the agent server, the communication model also accounts for the backhaul latency incurred during data transmission between them. This latency only arises in ground-to-ground RSU links, as satellite nodes typically possess sufficient on-orbit computing resources, eliminating the need for AVs connected to satellites to offload AI agent tasks to terrestrial RSUs. The backhaul latency consists of two components: the wired transmission latency and the relay latency, where the latter is influenced by processing, forwarding, and queuing delays along the relay path. Therefore, the backhaul latency is expressed as
\begin{equation}
    L_{m}^{\text{back}}(t)=\mathbb{I}(e_m^t=n)\left(\mathbb{I}(c_m^t\neq e_m^t)\frac{D_{m}^{\text{up}}(t)+D_{m}^{\text{down}}(t)}{R_N}+2\varphi_t^{\text{back}}d_{c_m^t\to e_m^t}\right).
\end{equation}
Here, $R_N$ is the wired transmission rate between RSUs for AI agent migrations, $\varphi_t^{\text{back}}$ is a positive coefficient of the backhaul latency, and $d_{c_m^t\to e_m^t}$ is the hop count between the connected edge server $c_m^t$ and the agent server $e_m^t$.

Finally, the total communication latency is calculated by
\begin{equation}
    L_{m}^{\text{comm}}(t)=L_{m}^{\text{up}}(t)+L_{m}^{\text{back}}(t)+L_{m}^{\text{down}}(t).
\end{equation}

\subsubsection{Migration model}
According to Section~\ref{discrete_action}, a moving AV $m$ executes a discrete action, namely privacy-aware AI agent migration, which introduces a migration latency. Let $S_m$ denote the AI agent model size of AV $m$. The incurred migration latency of AV $m$ between the previous agent server and the current one is then computed as
\begin{equation}
    L_{m}^{\text{mig}}(t) = \begin{cases}
        0, & e_m^t = e_m^{t-1}, \\[4pt]
        \frac{S_m}{R_N}+\varphi^{\text{mig}}d_{e_m^{t-1}\to e_m^t}, & e_m^t = n, e_m^{t-1} \in (\mathbb{N} \setminus \{n\}), \\[4pt]
        \frac{S_m}{R_{\text{sat}}^{\text{up}}}, 
        & e_m^t = S, e_m^{t-1} \in \mathbb{N}, \\[4pt]
        \frac{S_m}{R_{\text{sat}}^{\text{down}}}, 
        & e_m^t \in \mathbb{N}, e_m^{t-1} = S, \\[4pt]
    \end{cases}
\end{equation}
where $\varphi^{\text{mig}}$ denotes the RSU-to-RSU AI migration latency coefficient.
\subsubsection{Computation model}
In addition to wireless communication and migration operations, the process of delivering metaverse services to AVs by AI agents also incurs computation latency at the agent server. Specifically, the agent server $e_m^t$ is responsible for AI agent updates and task processing for the agentic AI-driven AV $m$, resulting in the computation latency as
\begin{equation}
    L_{m}^{\text{comp}}(t)=\mathbb{I}(e_m^t\in\mathbb{N})\frac{c_{m}D_{m}^{\text{task}}(t)}{f_{n}}+\mathbb{I}(e_m^t=S)\frac{c_{m}D_{m}^{\text{task}}(t)}{f_{\text{sat}}}.
\end{equation}
Here, $c_{m}$ is the number of Central Processing Unit (CPU) cycles to process per unit data uploaded from AV $m$ (in CPU cycles/bit), and $D_{m}^{\text{task}}(t)=\iota D_{m}^{\text{up}}(t)$ is the size of the requested AI agent task created based on the upload data with $\iota$ denoting the conversion factor. $f_n$ and $f_{\text{sat}}$ are the CPU frequencies of RSU $n$ and satellite $S$, respectively~\cite{wu2025security}.

Finally, by integrating the communication latency, migration latency, and computation latency, the total service response latency experienced by AV $m$ in time slot $t$ is expressed as
\begin{equation}
    L_{m}^t=L_{m}^{\text{comm}}(t)+L_{m}^{\text{mig}}(t)+L_{m}^{\text{comp}}(t).
\end{equation}

\subsection{Penalties for QoS loss}
QoS is another critical requirement in 6G-enabled vehicular metaverses~\cite{cui2025overview}. When agentic AI-driven AVs travel and request location-sensitive LBS, injecting excessive noise into their status messages can degrade QoS, leading to poor metaverse experiences such as inaccurate scene rendering and improper site recommendations. To prevent such situations, a penalty term for QoS loss regarding the distance between the actual and perturbed locations is introduced to constrain large-scale continuous actions. According to Eq. (\ref{perturbed_location}), the perturbed location of AV $m$ after applying the continuous action is $\tilde{l}_m^t=(\tilde{x}_m^t,\tilde{y}_m^t)$. Here, a logarithmic structure is adopted to capture the diminishing marginal impact of location perturbation on QoS degradation~\cite{datar2022strategic}. Specifically, small perturbations can already significantly affect location-sensitive metaverse services, whereas further increasing a large perturbation distance results in a relatively smoother QoS degradation. Therefore, the logarithmic penalty effectively models the progressively saturated QoS loss and prevents the QoS term from dominating the overall optimization objective. Referring to the distance calculation formula in Eq. (\ref{distance_formula}), we define the penalty for QoS loss as
\begin{equation}
    Q_m^t=\ln{\left(1 + 2R\arcsin{\sqrt{\sin^2 \left(\frac{\pi}{360}(x_m^t-\tilde{x}_m^t)\right) +\cos(\frac{\pi}{180}x_m^t)\cos(\frac{\pi}{180} \tilde{x}_m^t) \sin^2 \left((\frac{\pi}{360}(y_m^t-\tilde{y}_m^t)\right)}}\right)}.
\end{equation}

\subsection{Problem formulation}
\label{optimization_problem}
Generally, the proposed framework requires each agentic AI-driven AV to determine how to perturb its physical location and migrate its AI agent to protect cross-reality location privacy and ensure user immersion by maintaining service response latency and QoS. Consequently, the utility function of AV $m$ is defined as
\begin{equation}
    u_m^t = \omega_E E_m^t - \omega_L L_{m}^t - \omega_Q Q_{m}^t,
\end{equation}
where $\omega_E$, $\omega_L$, and $\omega_Q$ represent the weighting factors for the cross-reality location entropy, service response latency, and QoS loss, respectively. Finally, the overall optimization problem of the proposed hybrid action mechanisms in the 6G-enabled vehicular metaverse is formulated as
\begin{subequations}
\begin{align}
    \textbf{P1}:\max_{r_m^t,\theta_m^t,e_m^t}\:\: &\frac{1}{T} \frac{1}{M} \sum_{t=1}^{T} \sum_{m=1}^{M} u_m^t, \label{eq18a}\\
    \text {s.t.}\:\:&r_m^t \in [0, r_{\max}], \label{eq18b}\\
    \:\:&\theta_m^t \in [0, 2\pi), \label{eq18c}\\
    \:\:&\mathbb{I}(e_m^t\in\mathbb{N})+\mathbb{I}(e_m^t=S)=1. \label{eq18d}
\end{align}
\end{subequations}
Here, $r_{\max}$ is the maximum perturbation radius of AVs. Constraint (\ref{eq18b}) limits the perturbation radius to a feasible service area, preventing excessively large location deviations that would severely degrade LBS quality. Constraint (\ref{eq18c}) specifies the complete angular range for two-dimensional location perturbation, allowing the AV to choose any perturbation direction in the physical plane. Constraint (\ref{eq18d}) enforces a valid virtual-space migration decision by ensuring that each AI agent is hosted by exactly one candidate edge server, either an RSU or the LEO satellite, at each time slot. It is observed that the formulated problem $\textbf{P1}$ is a complex non-convex mixed-integer programming problem. Since cross-reality location privacy preservation is persistent and varies over time, traditional convex optimization methods that provide only single-round static solutions are often invalid. Moreover, the long-term coupling among consecutive decisions in dynamic vehicular metaverse scenarios makes learning-based methods more suitable than conventional optimization for achieving sustained privacy and immersion performance.

\section{LLM-enhanced hybrid GDM-based reinforcement learning solution}
\label{DRL_diffusion}
Although it is difficult to solve the optimization problem directly, the problem $\textbf{P1}$ exhibits a memoryless sequential decision-making structure that satisfies the Markov property. Based on this observation, we model the hybrid action decision process as a Markov Decision Process (MDP)~\cite{wang2023location}. Moreover, large AI models exhibit strong capabilities to understand contextual information in highly dynamic systems, among which LLMs are particularly effective at reasoning over abstract objectives and discovering latent performance indicators~\cite{zhang2025personalizing,li2026carbongpt}. Motivated by recent advances in LLM-enhanced reinforcement learning~\cite{cao2024survey,zhang2026toward}, we propose an LLM-enhanced Hybrid Diffusion-based Proximal Policy Optimization (LHDPPO) algorithm to solve the complex optimization problem $\textbf{P1}$.

\subsection{Markov decision process modeling}
\subsubsection{Environment state} Let $\check{l}_m^t$ denote the position of the connected server $c_m^t$. Thus, the state space of AV $m$ in time slot $t \in \{1,2,\ldots,T\}$ is defined as a union of the real-time vehicle location $l_m^t$, the location of the connected edge server $\check{l}_m^t$, and the location of the edge server hosting the AI agent in the previous slot $\hat{l}_m^{t-1}$, which is represented by
	\begin{equation}
		s_m^t \triangleq \{l_m^t, \dot{\theta}_m^t, \check{l}_m^t,    \hat{l}_m^{t-1}\},
	\end{equation}
where $\dot{\theta}_m^t \in [0,\pi]$ denotes the yaw angle of AV $m$ relative to the road network direction in time slot $t$. Therefore, the state space in the MDP is the aggregation of the states of all AVs, denoted as $\boldsymbol{s}^t=\{s_m^t\}_{m=1}^M$.

\subsubsection{Action space} At each time slot $t$, each agentic AI-driven AV autonomously determines a hybrid action consisting of a continuous action $\boldsymbol{c}_m^t$, comprising the perturbation radius $r_m^t$ and perturbation angle $\theta_m^t$ in reality, and a discrete action $\boldsymbol{d}_m^t$, i.e., the AI agent migration decision $e_m^t$ in virtuality. Therefore, the action of AV $m$ is denoted as $\boldsymbol{a}_m^t\triangleq \{\boldsymbol{c}_m^t,\boldsymbol{d}_m^t\}= \{r_m^t,\theta_m^t,e_m^t\}$. Consequently, the action space in the MDP is defined as $\boldsymbol{a}^t=\{\boldsymbol{c}^t,\boldsymbol{d}^t\}=\{\boldsymbol{a}_m^t\}_{m=1}^{M}$.

\subsubsection{Reward function}
The reward function characterizes the immediate feedback received by AVs after executing the hybrid action $\boldsymbol{a}^t$ given the environment state $\boldsymbol{s}^t$, which plays a critical role in policy updates within the MDP. Rather than directly defining the reward as the AV utility, a trajectory-consistency penalty is incorporated to constrain the directional deviation between perturbed trajectories and their corresponding real trajectories. This penalty encourages the perturbed trajectories to better conform to realistic road-network mobility patterns, thereby improving the authenticity of the generated perturbations. Therefore, the reward function is defined as
\begin{equation}
\mathcal{R}_{\mathrm{man}}(\boldsymbol{s}^t,\boldsymbol{a}^t)
=
\frac{1}{M}\sum_{m=1}^{M}
\left(
u_m^t-\omega_{\Phi}[\Phi_m^t-\Phi_{\max}]^{+}
\right),
\end{equation}
Here, $\Phi_m^t = \sin{\left[\min\left( \left| \theta_m^t - \dot{\theta}_m^t \right|,\; 2\pi - \left| \theta_m^t - \dot{\theta}_m^t \right| \right)\right]}$ is the direction inconsistency score. Meanwhile, $\Phi_{\max}$ is a tolerable direction inconsistency threshold and $[x]^+ \triangleq \max\{x,0\}$. However, such manually crafted rewards are often limited to predefined objective terms and penalty components, and may fail to capture implicit interactions among location privacy, service response latency, and QoS in the complex cross-reality vehicular metaverse scenario, potentially leading to suboptimal convergence.

\begin{figure*}[t]
    \centerline{\includegraphics[width=0.9\textwidth]{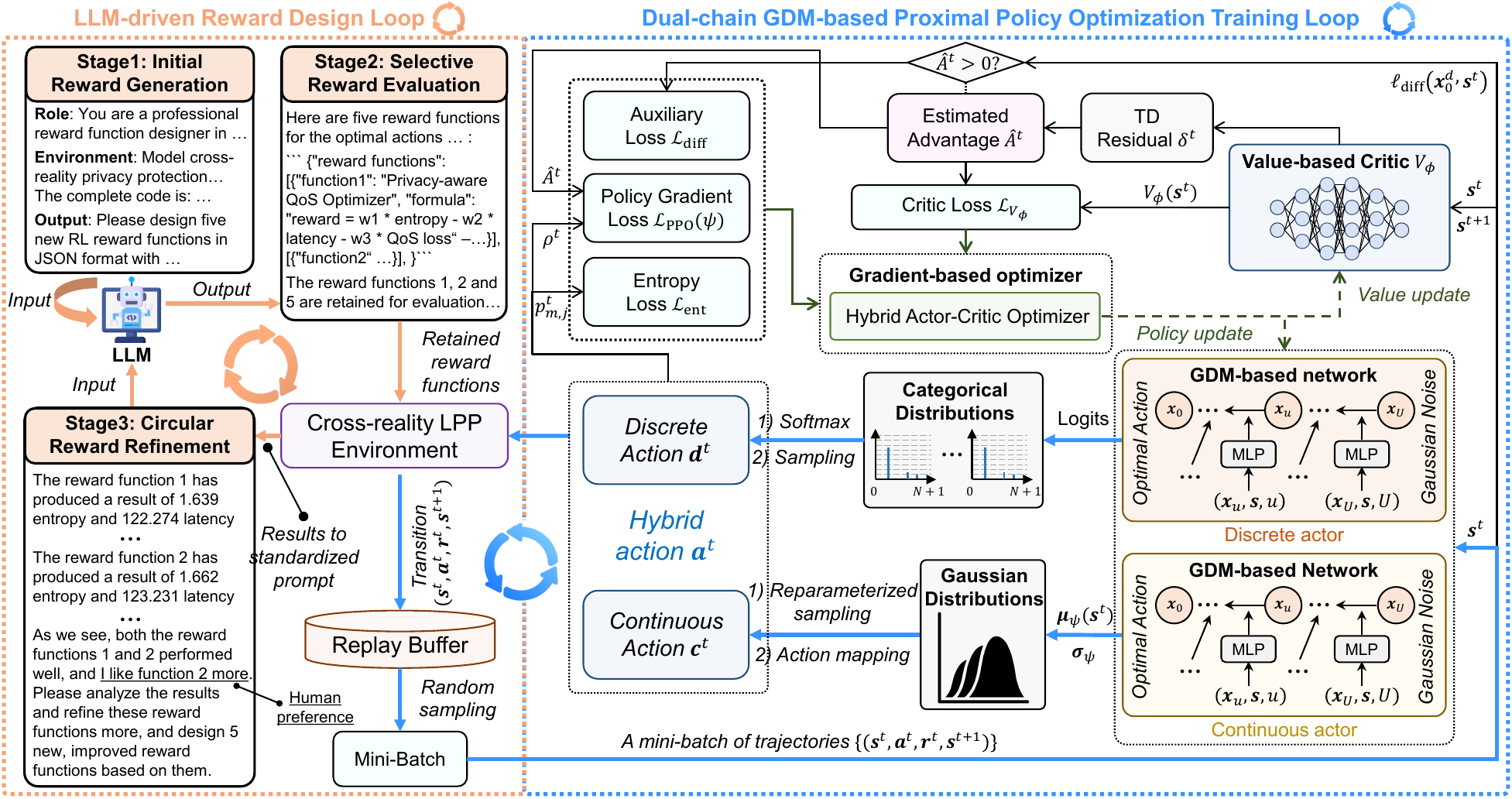}}
    \caption{(Color online) Architecture of the proposed LHDPPO algorithm for the cross-reality location privacy protection (LPP) environment.}
    \label{fig2}
\end{figure*}

\subsection{LHDPPO algorithm details}
\subsubsection{LLM-driven reward function design loop}
Instead of relying solely on manual formulations that may neglect the important environmental components, the proposed LHDPPO algorithm employs an LLM as an autonomous reward designer to generate more informative reward functions. This design philosophy is consistent with the role of agentic AI, where the LLM is integrated into the reinforcement learning loop to enhance policy learning through adaptive reward shaping.

As illustrated in Figure~\ref{fig2}, the LLM-driven reward design follows an iterative loop consisting of three main stages.

\textbf{Stage 1: Initial reward generation}: The LLM is first prompted with three categories of inputs to produce a set of candidate reward functions. These inputs include
\begin{itemize}
    \item Role assignment: The prompt instructs the LLM to act as an experienced and professional reward designer with expertise in wireless communications, security and privacy, and deep reinforcement learning.
    \item Environment description: The prompt provides a detailed description of the 6G-enabled vehicular metaverse scenario together with the complete environment code, including the MDP formulation and parameter configuration.
    \item Output specification: The prompt requires the LLM to generate reward functions in a structured JSON format, accompanied by executable Python code that can be directly integrated into the simulation environment.
\end{itemize}

\textbf{Stage 2: Selective reward evaluation}: Based on the generated candidate reward functions, domain experts selectively retain those that are suitable for training and discard unreasonable ones, such as functions that assign excessive weight to a single term. Each retained reward function is then evaluated through interactions with the environment, and key performance indicators such as achieved location entropy and service latency are collected.

\textbf{Stage 3: Circular reward refinement}: The evaluation results from Stage 2 are standardized and fed back to the LLM as refinement prompts. Guided by explicit performance comparisons and optional human preferences, the LLM iteratively updates and improves the reward formulations by emphasizing influential components that are often overlooked in manual designs, such as the coupled effects of privacy, latency, and QoS. This closed reward design loop continues until convergence to a high-performing reward function, as shown on the left side of Figure~\ref{fig2}.

After multiple refinement rounds, the LLM outputs the most informative reward function for cross-reality location privacy protection, which is finally adopted by the learning agent:
\begin{equation}
\boldsymbol{r}^t =\mathcal{R}_{\text{LLM}}(\boldsymbol{s}^t,\boldsymbol{a}^t)= \frac{1}{M}\sum_{m=1}^M \left(\omega_1 E_m^t - \omega_2 \ln{(1+L_{m}^t)} - \omega_3 \hat{Q}_m^t -\omega_4 [\Phi_m^t-\Phi_{\max}]^{+} -\omega_5 \ln{(1+d_{m,e_m^t}^t)} + \omega_6 \alpha_m^t\right).
\end{equation}
Here, $\hat{Q}_m^t=\frac{Q_m^t}{Q_{\max}}$ is the normalized QoS loss, with $Q_{\max}$ representing the maximum QoS loss under the maximum perturbation radius $r_{\max}$. $d_{m,e_m^t}^t$ is the distance between the AV $m$ and the selected agent server $e_m^t$. $\alpha_m^t$ is the included angle between the vehicle-to-server direction and perturbation direction. $\omega_1$, $\omega_2$, $\omega_3$, $\omega_4$, $\omega_5$, and $\omega_6$ are the optimized weighting factors obtained through multiple rounds of LLM-guided reward refinement, which yield superior learning performance. Compared with $\mathcal{R}_{\text{man}}(\cdot)$, the LLM-generated reward incorporates latent yet critical indicators and explicitly balances location privacy with user immersion, enabling the proposed LHDPPO algorithm to better improve policy robustness and convergence performance in 6G-enabled vehicular metaverses.

\subsubsection{Dual-chain GDM-based PPO training loop}
Beyond the LLM-driven reward design, the proposed LHDPPO algorithm integrates powerful GDMs that can model complex, high-dimensional distributions through multi-step denoising processes, enabling structured action exploration in the highly coupled decision space of 6G-enabled vehicular metaverses. Specifically, LHDPPO leverages two diffusion policies to handle continuous and discrete actions, respectively, jointly forming the hybrid policy as
\begin{equation}
\pi_{\psi}(\boldsymbol{a}^t|\boldsymbol{s}^t)
=\pi_{\psi^{c}}^{c}(\boldsymbol{c}|\boldsymbol{s})\,
\pi_{\psi^{d}}^{d}(\boldsymbol{d}|\boldsymbol{s}),
\end{equation}
where $\boldsymbol{c}=\{r_m,\theta_m\}_{m=1}^M$ denotes the continuous actions for location perturbation and $\boldsymbol{d}=\{e_m\}_{m=1}^M$ denotes the discrete actions for privacy-aware AI agent migration. The key idea of GDM-based algorithms is to exploit the reverse diffusion process to progressively refine actions from noise into feasible solutions, thereby enabling effective exploration for near-optimal hybrid actions. This generative denoising process is conceptually described as
\begin{equation}
p_{\psi}(\boldsymbol{x}_{0:U}|\boldsymbol{s})
=\mathcal{N}(\boldsymbol{x}_U;\boldsymbol{0},\mathbf{I})
\prod_{u=1}^{U} p_{\psi}(\boldsymbol{x}_{u-1}|\boldsymbol{x}_u,\boldsymbol{s}),
\end{equation}
where $U$ is the number of denoising steps, $\mathbf{I}$ is the covariance matrix, and $\boldsymbol{x}_0$ is the final output of the reverse chain.

The reverse diffusion chain $\{p_{\psi}(\boldsymbol{x}_{u-1}|\boldsymbol{x}_u,\boldsymbol{s})\}_{u=1}^{U}$ is modeled as a sequence of conditional Gaussian transitions,
\begin{equation}
p_{\psi}(\boldsymbol{x}_{u-1}|\boldsymbol{x}_u,\boldsymbol{s})
= \mathcal{N}\!\left(\boldsymbol{x}_{u-1};
\boldsymbol{\mu}_{\psi}(\boldsymbol{x}_u,\boldsymbol{s},u),
\boldsymbol{\Sigma}_{\psi}(\boldsymbol{x}_u,\boldsymbol{s},u)\right),
\end{equation}
where the covariance follows a predefined variance schedule $\boldsymbol{\Sigma}_{\psi}(\boldsymbol{x}_u,\boldsymbol{s},u) = \beta_u \mathbf{I}$ and the mean is parameterized via the noise prediction network $\boldsymbol{\varepsilon}_{\psi}$ as
\begin{equation}
\boldsymbol{\mu}_{\psi}(\boldsymbol{x}_u,\boldsymbol{s},u)
= \frac{1}{\sqrt{\alpha_u}}
\left(
\boldsymbol{x}_u
- \frac{\beta_u}{\sqrt{1-\bar{\alpha}_u}}\,
\boldsymbol{\varepsilon}_{\psi}(\boldsymbol{x}_u,\boldsymbol{s},u)
\right),
\end{equation}
with $\alpha_u = 1 - \beta_u$ and $\bar{\alpha}_u = \prod_{k=1}^{u} \alpha_k$.

Accordingly, the hybrid policy $\pi_{\psi}(\boldsymbol{a}^t|\boldsymbol{s}^t)$ is realized by executing the reverse diffusion process from $\boldsymbol{x}_U \sim \mathcal{N}(\mathbf{0},\mathbf{I})$ to $\boldsymbol{x}_0$, where the final denoised sample $\boldsymbol{x}_0$ corresponds to the executed action $\boldsymbol{a}^t$. The same reverse diffusion principle is applied to both continuous and discrete components, yielding $\boldsymbol{c}$ and $\boldsymbol{d}$, respectively. Consequently, optimizing the policy $\pi_{\psi}$ in the high-dimensional vehicular metaverse environment is equivalent to optimizing the denoising networks $\boldsymbol{\varepsilon}_{\psi^{c}}^{c}$ and $\boldsymbol{\varepsilon}_{\psi^{d}}^{d}$ that govern the reverse diffusion dynamics of the continuous and discrete action spaces. These denoising networks fully determine the action distribution, which ultimately determines the behavior of the learning agent.

At time slot $t$, the hybrid log-probability of the executed action generated by the diffusion policy is computed as
\begin{equation}
\log \pi_{\psi}(\boldsymbol{a}^t|\boldsymbol{s}^t)
=\log \pi_{\psi^{c}}^{c}(\boldsymbol{c}^t|\boldsymbol{s}^t)
+\log \pi_{\psi^{d}}^{d}(\boldsymbol{d}^t|\boldsymbol{s}^t).
\end{equation}

In LHDPPO, both the continuous and discrete policies are implemented by diffusion-based actors that execute the reverse diffusion chain in a mean-only denoising mode. The stochasticity required for policy optimization is solely introduced by the final distributions, which ensures tractable likelihood evaluation and stable PPO updates.

For the continuous component, the diffusion actor produces a mean $\boldsymbol{\mu}_{\psi^{c}}(\boldsymbol{s}^t)$, and the continuous policy is then modeled as a Gaussian distribution, given by
\begin{equation} \pi_{\psi^{c}}^{c}
    (\boldsymbol{c}^t|\boldsymbol{s}^t) = \mathcal{N}\!\left(\boldsymbol{c}^t; \boldsymbol{\mu}_{\psi^{c}}(\boldsymbol{s}^t), \operatorname{diag}(\boldsymbol{\sigma}_{\psi^{c}}^2)\right), 
\end{equation} 
where the log-standard-deviation vector $\boldsymbol{\sigma}_{\psi^{c}}$ is learned jointly with the diffusion model parameters. This construction enables the PPO policy gradient to propagate through the log-probability of the sampled action into the continuous denoising network $\boldsymbol{\varepsilon}_{\psi^{c}}^{c}$, thereby shaping the reverse diffusion dynamics according to long-term system performance.

For the discrete component, a diffusion-based denoising network generates the categorical logits for each AV, from which the migration action $e_m^t$ is sampled. Then, the corresponding log-probability is computed by
\begin{equation}
\log \pi_{\psi^{d}}^{d}(\boldsymbol{d}^t|\boldsymbol{s}^t)
=\sum_{m=1}^{M} \log \text{Cat}(e_m^t \,|\, \text{logits}_m(\boldsymbol{s}^t)).
\end{equation}
Here, $\text{Cat}(\cdot)$ denotes the categorical distribution parameterized by $\text{logits}_m(\boldsymbol{s}^t)$. After applying the softmax operation to these logits, the probability of selecting each candidate edge server is obtained, and $\log \text{Cat}(e_m^t \,|\, \text{logits}_m(\boldsymbol{s}^t))$ represents the log-probability assigned to the actually sampled migration action $e_m^t$.
The gradients of this log-probability propagate through the diffusion-generated logits and update the discrete denoising network $\boldsymbol{\varepsilon}_{\psi^{d}}^{d}$, thereby completing the coupling between diffusion modeling and PPO-based policy optimization.

For PPO updates, we denote $\pi_{\psi_{\text{old}}}$ as the old hybrid behavior policy, and the importance ratio is computed as
\begin{equation}
\rho^t(\psi)
=\frac{\pi_{\psi}(\boldsymbol{a}^t|\boldsymbol{s}^t)}
{\pi_{\psi_{\text{old}}}(\boldsymbol{a}^t|\boldsymbol{s}^t)}
=\exp\!\left(
\log \pi_{\psi}(\boldsymbol{a}^t|\boldsymbol{s}^t)
-\log \pi_{\psi_{\text{old}}}(\boldsymbol{a}^t|\boldsymbol{s}^t)
\right).
\end{equation}

In LHDPPO, a learned value-based critic network $V_{\phi}(\boldsymbol{s})$ is employed to approximate the expected value of the current state, with parameters $\phi$ updated from sampled trajectories. Based on this approximation, the Generalized Advantage Estimation (GAE) is constructed as $\hat{A}^t = \sum_{l=0}^{\infty} (\gamma \lambda)^l \delta^{t+l}$, where $\lambda$ is the GAE parameter, $\gamma$ is the discount factor, and $\delta^t = \boldsymbol{r}^t + \gamma V_{\phi}(\boldsymbol{s}^{t+1}) - V_{\phi}(\boldsymbol{s}^t)$ is the Temporal-Difference (TD) residual. This advantage signal can drive the policy update by weighting the likelihood ratio in the PPO clipped surrogate objective, given by~\cite{zhang2025embodied}
\begin{equation}
\mathcal{L}_{\text{PPO}}(\psi)
= - \mathbb{E}^t
\left[
\min\!\left(
\rho^t(\psi)\hat{A}^t,\;
\text{clip}(\rho^t(\psi),1-\kappa,1+\kappa)\hat{A}^t
\right)
\right],
\end{equation}
where $\kappa\in[0,1]$ is the clipping parameter. The gradient of this objective is backpropagated through the log-probability of the diffusion policy and ultimately updates the denoising networks $\boldsymbol{\varepsilon}_{\psi^{c}}^{c}$ and $\boldsymbol{\varepsilon}_{\psi^{d}}^{d}$. In this manner, LHDPPO forms another closed training loop in which GDM-based action generation, environment interaction, reward evaluation, and PPO updates are tightly coupled, as shown on the right side of Figure~\ref{fig2}.

Meanwhile, the value network is trained using the clipped value loss, given by
\begin{equation}
\mathcal{L}_{V_\phi}
= \frac{1}{2}\,
\mathbb{E}^t
\left[
\max\!\left(
(V_{\phi}(\boldsymbol{s}^t)-\hat{R}^t)^2,\;
(V_{\text{clip}}(\boldsymbol{s}^t)-\hat{R}^t)^2
\right)
\right].
\end{equation}
Here, $\hat{R}^t = \hat{A}^t + V_{\phi}(\boldsymbol{s}^t)$ is the return target for value learning. $V_{\text{clip}}(\boldsymbol{s}^t)
= V_{\text{old}}(\boldsymbol{s}^t)
+ \text{clip}\!\left(
V_{\phi}(\boldsymbol{s}^t)-V_{\text{old}}(\boldsymbol{s}^t),
-\kappa,\kappa
\right)$ is the clipped value prediction used to prevent excessively large value updates, with $V_{\text{old}}(\boldsymbol{s}^t)$ denoting the value estimate stored when the trajectory was collected.

To prevent premature convergence of the discrete migration policy in the extremely large action space, LHDPPO introduces an entropy regularization loss for the discrete categorical distributions, calculated by
\begin{equation}
\mathcal{L}_{\text{ent}}
= - c_{\text{ent}}^t\left[\frac{1}{M}
\sum_{m=1}^{M}
\left(
- \sum_{j=1}^{N+1} p_{m,j}^t \log p_{m,j}^t
\right)\right].
\end{equation}
Here, $p_{m,j}^t = \Pr(e_m^t = j \mid \boldsymbol{s}^t)$ denotes the probability assigned by the discrete diffusion policy to migrating the AI agent of AV $m$ to candidate destination edge server $j$ at time slot $t$, with $j \in \mathbb{N} \cup \{S\}$. $c_{\text{ent}}^t$ is an annealed entropy coefficient to ensure stable exploration and efficient convergence during training.

Meanwhile, to stabilize learning of the discrete diffusion actor in the large and multi-dimensional discrete action space, an auxiliary diffusion denoising loss is introduced. The executed migration decisions are converted into a concatenated one-hot vector $\boldsymbol{x}_0^d \in \{0,1\}^{M(N+1)}$ and optimized through a diffusion reconstruction objective. This auxiliary loss is applied only to samples with a positive estimated advantage to avoid reinforcing suboptimal behaviors, and is defined as
\begin{equation}
\mathcal{L}_{\text{diff}}
= \mathbb{E}\!\left[ \mathbb{I}(\hat{A}^t>0)\, \ell_{\text{diff}}(\boldsymbol{x}_0^d,\boldsymbol{s}^t) \right],
\end{equation}
where $\ell_{\text{diff}}(\boldsymbol{x}_0^d,\boldsymbol{s}^t)$ measures the reconstruction error of the discrete diffusion model when denoising the executed migration vector conditioned on state $\boldsymbol{s}^t$ and is instantiated as an L1 reconstruction loss in our implementation.

Eventually, the overall training objective of the proposed LHDPPO algorithm is given by
\begin{equation}
\mathcal{L}
= \mathcal{L}_{\text{PPO}}
+ c_v \mathcal{L}_{V_\phi}
+ \mathcal{L}_{\text{ent}}
+ c_{\text{aux}}^t \mathcal{L}_{\text{diff}},
\end{equation}
where $c_v$ and $c_{\text{aux}}^t$ denote the value loss coefficient and the annealed auxiliary diffusion coefficient, respectively.

Complexity analysis: Let $\chi\in\{c,d,v\}$ denote the continuous denoising actor, discrete denoising actor, and value critic, respectively. For $\chi$, let $H_\chi$ and $Z_h^\chi$ denote the number of layers and neurons in the $h$-th layer. Hence, the forward cost is approximated as $F_\chi=\sum_{h=1}^{H_\chi-1}(2Z_h^\chi-1)Z_{h+1}^\chi$~\cite{wang2025uplink}. Since LHDPPO employs two diffusion chains, one action generation step requires $U(F_c+F_d)$ operations for $U$ denoising steps. With $K$ PPO epochs, mini-batch size $B$, and trajectory length $T$, the training complexity is $\mathcal{O}\!\left(KBT[U(F_c+F_d)+F_c+F_d+F_v]\right)$, while online inference only needs $\mathcal{O}\!\left(U(F_c+F_d)\right)$ per decision slot. This indicates that the main extra cost of LHDPPO comes from the denoising steps and network widths, while the two diffusion chains can be evaluated in parallel.

\begin{figure*}[t]
    \centerline{\includegraphics[width=0.95\textwidth]{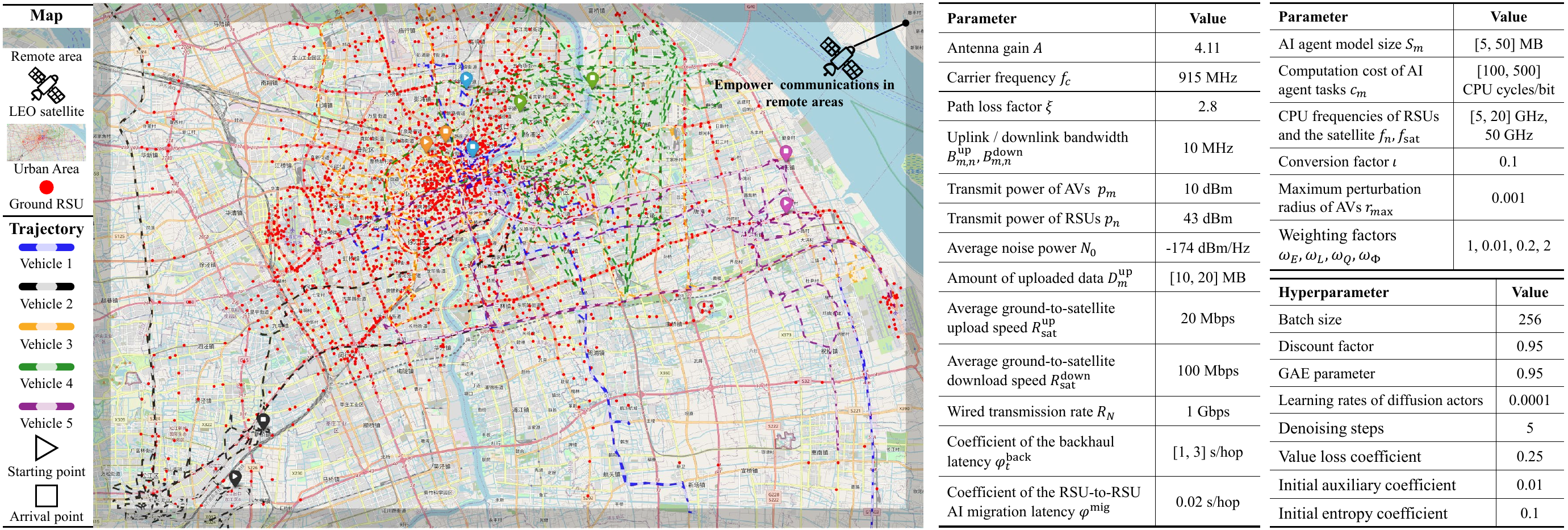}}
    \caption{(Color online) Left: Visualization of AV trajectories and edge server locations. Right: Default parameters and hyperparameters.}
    \label{fig3}
\end{figure*}

\section{Experimental results}
\label{Performance Evaluation}
\subsection{Experimental settings}
As illustrated on the left side of Figure~\ref{fig3}, we use 24-hour mobility traces of five taxis selected from a Shanghai dataset containing 4,316 taxis on February 20, 2007~\cite{zhu2010impact}. RSUs are extracted from a real-world dataset of about 840,000 base stations across China\footnote{https://gitcode.com/open-source-toolkit/abc64} and filtered within the target urban region, while an LEO satellite is incorporated to provide ubiquitous 6G connectivity in remote areas at the boundary of the map. Following~\cite{wang2023location,chen2024distributed,luo2025incentivizing}, the environmental parameters and hyperparameters adopted in the experiments are presented on the right side of Figure~\ref{fig3}. Moreover, LHDPPO is compared with the following five representative baseline methods.

\begin{itemize}
    \item \textbf{Geo-Indistinguishability (Geo-I)}~\cite{andres2013geo}: A non-learning method that perturbs location based on differential privacy, and employs a greedy strategy to select the agent server with the best historical performance.
    \item \textbf{Hybrid PPO (H-PPO)}~\cite{fan2019hybrid}: A classical hybrid reinforcement learning algorithm in which discrete action selection and continuous parameter optimization are trained in parallel using PPO.
    \item \textbf{GDMs-assisted Hybrid PPO (GHPPO)}~\cite{wang2025uplink}: A hybrid PPO method that introduces GDMs into parameterized action generation for mixed discrete-continuous decision optimization.
    \item \textbf{Hybrid-GDM (HGDM)}~\cite{kang2025hybrid}: A GDM-based Soft Actor-Critic (SAC) algorithm that expresses both discrete and continuous decisions as continuous latent variables and maps them to executable hybrid actions during inference.
    \item \textbf{PPO-relaxed}~\cite{zhang2025embodied}: A PPO-based algorithm employing clipped surrogate objectives and GAE for stable policy updates, where discrete migration actions are similarly relaxed into continuous values for tractable training.
\end{itemize}

\subsection{Numerical results}
\begin{figure*}[t]
    \centering
    \subfloat[\label{fig4:a}]{
    \includegraphics[width=0.31\textwidth]{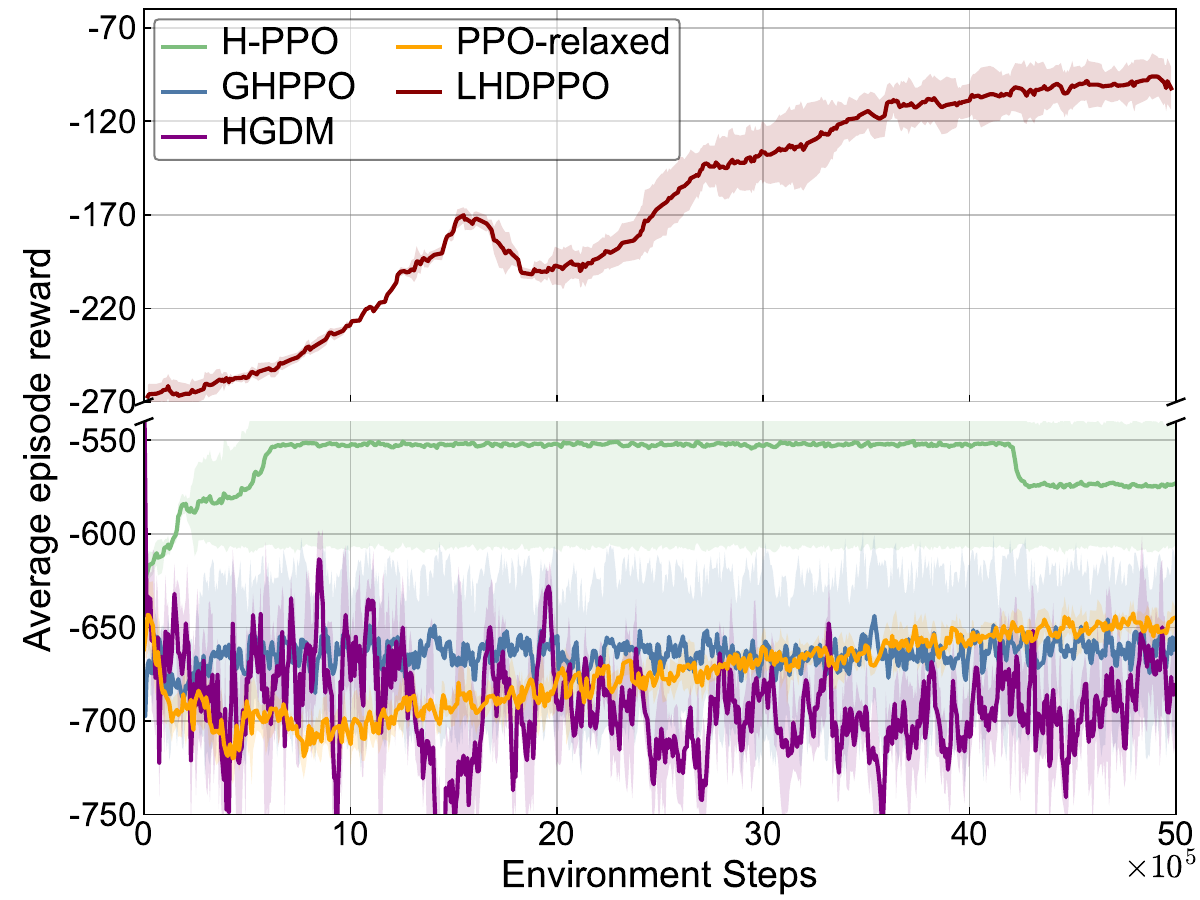}}
    \hfill
    \subfloat[\label{fig4:b}]{
    \includegraphics[width=0.32\textwidth]{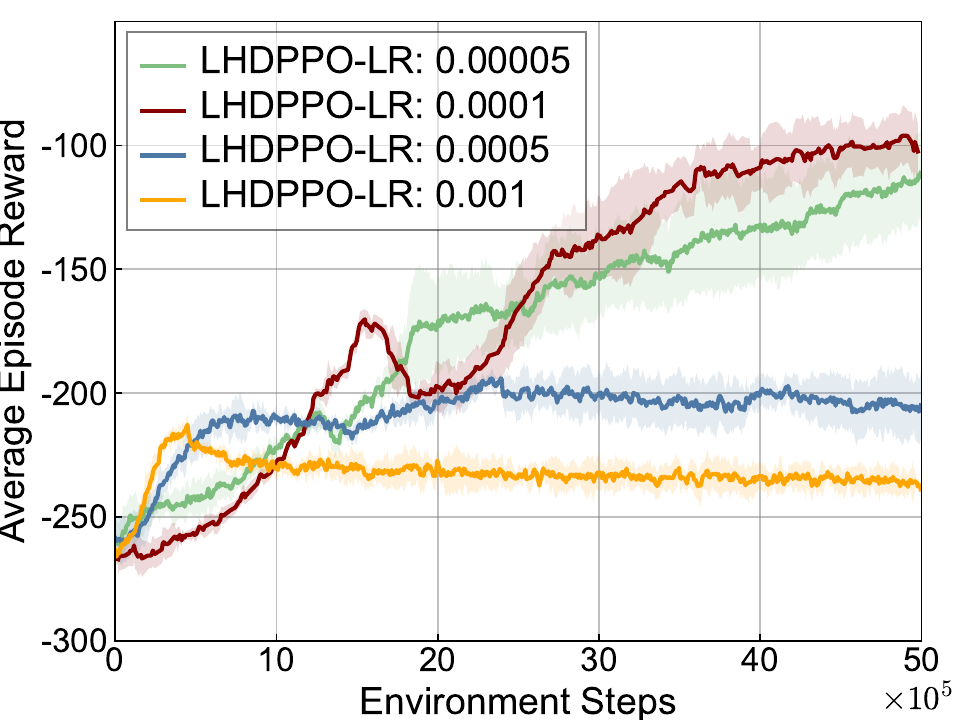}}
    \hfill
    \subfloat[\label{fig4:c}]{
    \includegraphics[width=0.32\textwidth]{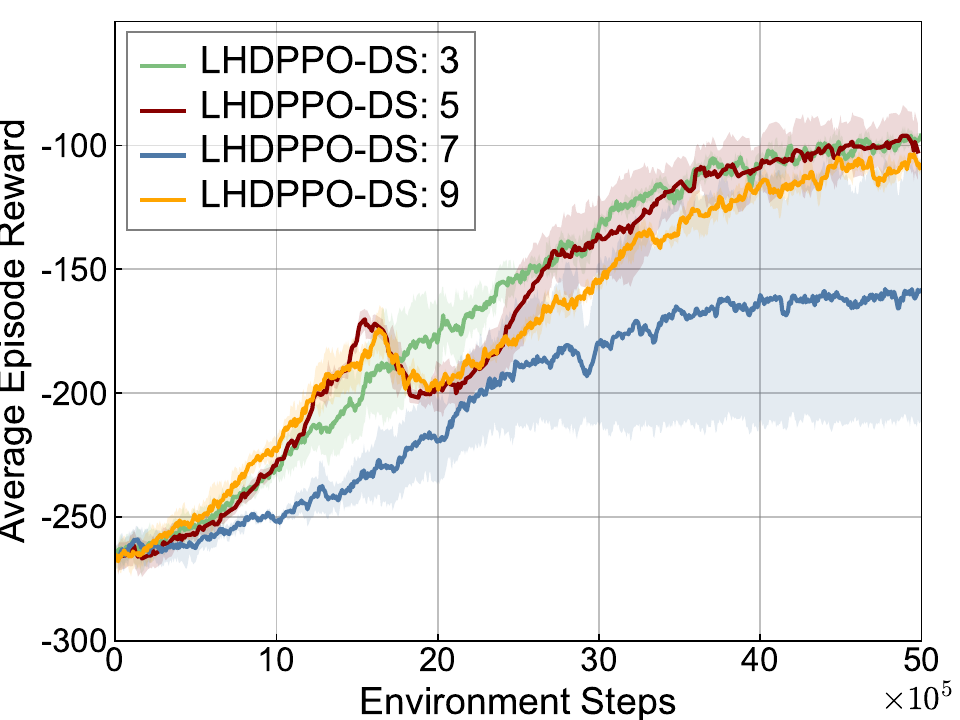}}
    \caption{(Color online) Convergence performance of the proposed LHDPPO algorithm and the baseline methods. Curves and shaded regions represent the mean $\pm$ standard deviation over five independent runs under different random seeds. LR: learning rate. DS: denoising step. (a) Test reward curves of LHDPPO and baseline methods for hybrid action decisions; (b) Test reward curves of LHDPPO under different learning rate settings; (c) Test reward curves of LHDPPO under different denoising step settings.}
    \label{fig4}
\end{figure*}

\textbf{Convergence analysis.}
As shown in Figure~\ref{fig4:a}, the proposed LHDPPO achieves a higher and more stable reward than the baselines, whereas several baselines show faster short-term reward gains but then saturate at lower performance levels or suffer from persistent oscillations, indicating difficulty in reliably optimizing the hybrid action mechanism. Figure~\ref{fig4:b} shows that the moderate learning rate of $10^{-4}$ improves the final convergence reward by 57.4\% over $10^{-3}$, whereas larger or smaller learning rates cause unstable or slow updates. This demonstrates that the use of a moderate learning rate can ensure reliable policy refinement. Figure~\ref{fig4:c} evaluates the influence of the number of diffusion denoising steps. Compared with 7 denoising steps, the default setting of 5 steps improves the final convergence reward by approximately 37.4\%, demonstrating that an intermediate denoising-step setting better balances convergence stability and final reward.

\begin{figure*}[t]
    \centering
    \subfloat[\label{fig:tb_utility}]{
        \includegraphics[width=0.235\textwidth]{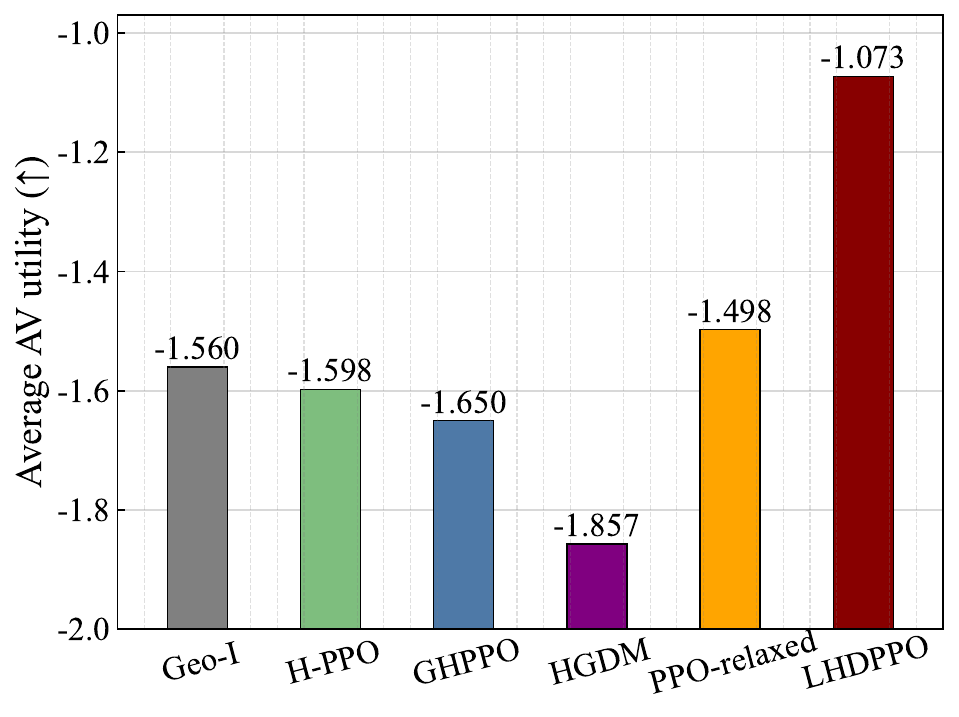}
    }\hfill
    \subfloat[\label{fig:tb_entropy}]{
        \includegraphics[width=0.235\textwidth]{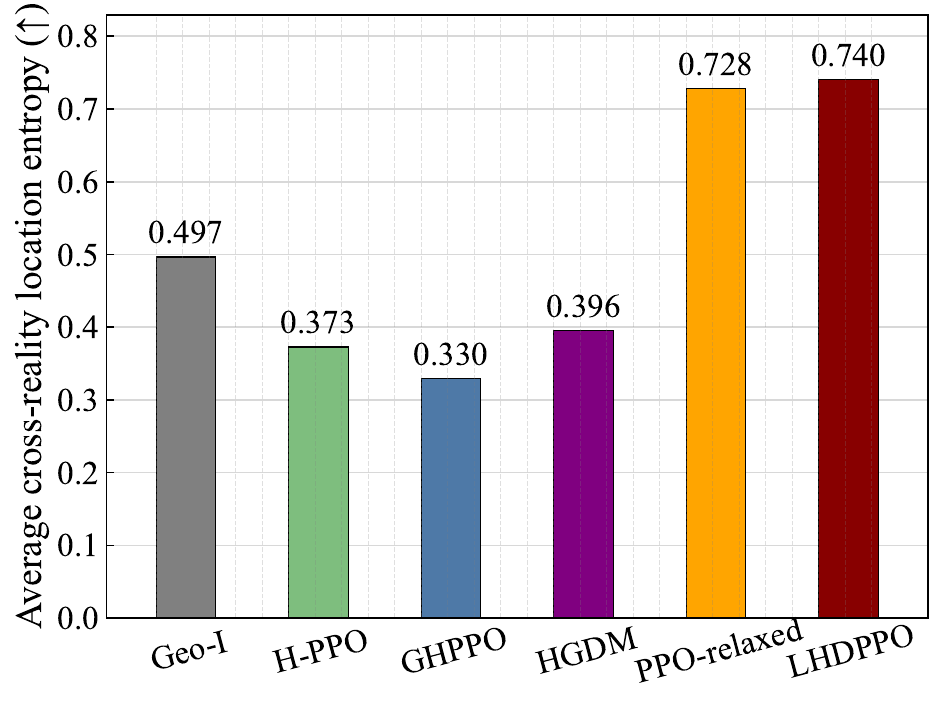}
    }\hfill
    \subfloat[\label{fig:tb_delay}]{
        \includegraphics[width=0.235\textwidth]{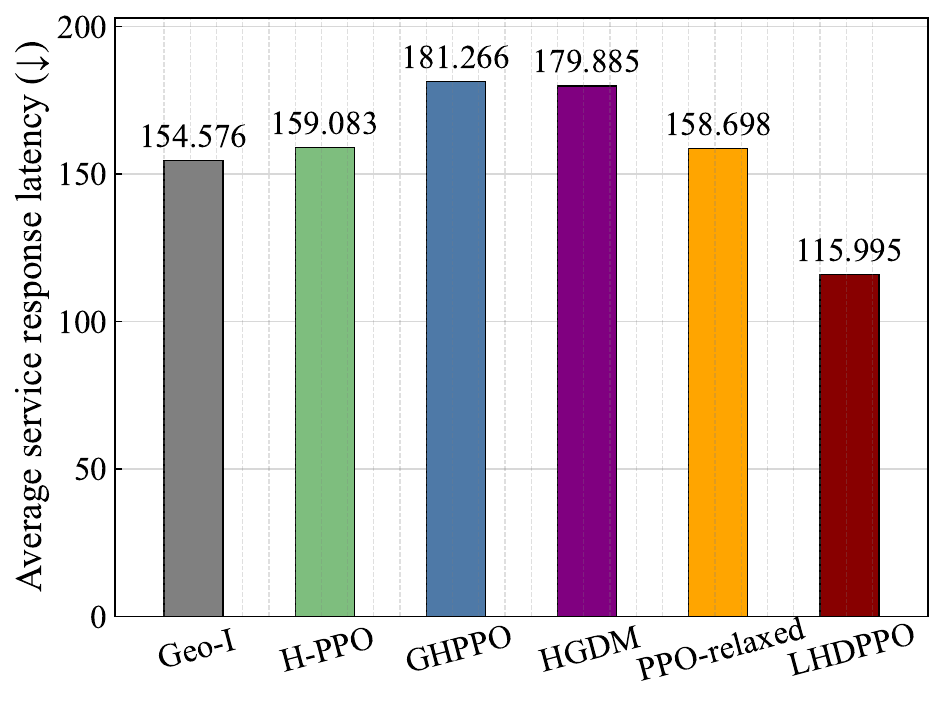}
    }\hfill
    \subfloat[\label{fig:tb_qos}]{
        \includegraphics[width=0.235\textwidth]{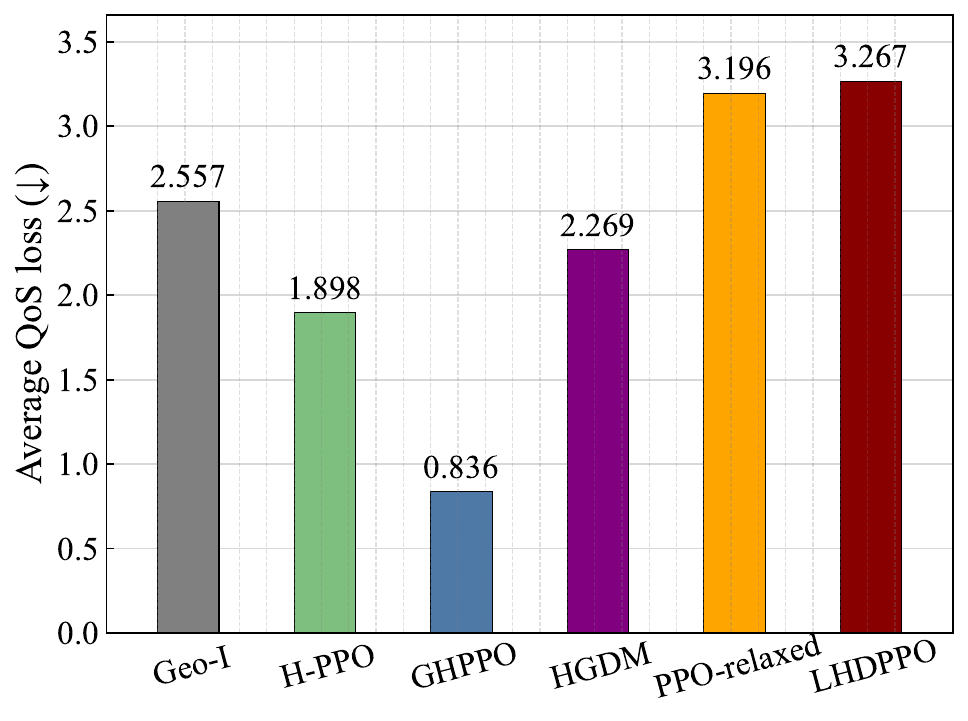}
    }

    \caption{(Color online) Performance comparison between the proposed LHDPPO algorithm and the baseline methods. (a) AV utility; (b) Cross-reality location entropy; (c) Service response latency; (d) QoS loss.}
    \label{fig5}
\end{figure*}

\textbf{Performance analysis.} Figure~\ref{fig5} compares the averaged test performance of LHDPPO and the representative baselines under the cross-reality location privacy protection framework. As shown in Figure~\ref{fig:tb_utility}, LHDPPO achieves the best AV utility of $-1.073$ and improves AV utility by 28.4\% compared with the strongest PPO-relaxed baseline method, demonstrating that the proposed hybrid action learning can better balance privacy protection, latency reduction, and QoS maintenance. As shown in Figure~\ref{fig:tb_entropy}, LHDPPO also obtains the highest cross-reality location entropy of 0.740, which improves the entropy by 48.9\% over Geo-I, indicating that the joint optimization of physical perturbation and virtual AI agent migration can more effectively resist cross-reality inference attacks. From Figure~\ref{fig:tb_delay} and Figure~\ref{fig:tb_qos}, we observe that LHDPPO reduces the service response latency by 26.9\% compared with PPO-relaxed and by 36.0\% compared with GHPPO, while its slightly higher QoS loss remains acceptable because the overall utility already reflects the combined privacy-latency-QoS tradeoff.

\begin{figure*}[t]
    \centering
    \subfloat[\label{fig6:a}]{
    \includegraphics[width=0.32\textwidth]{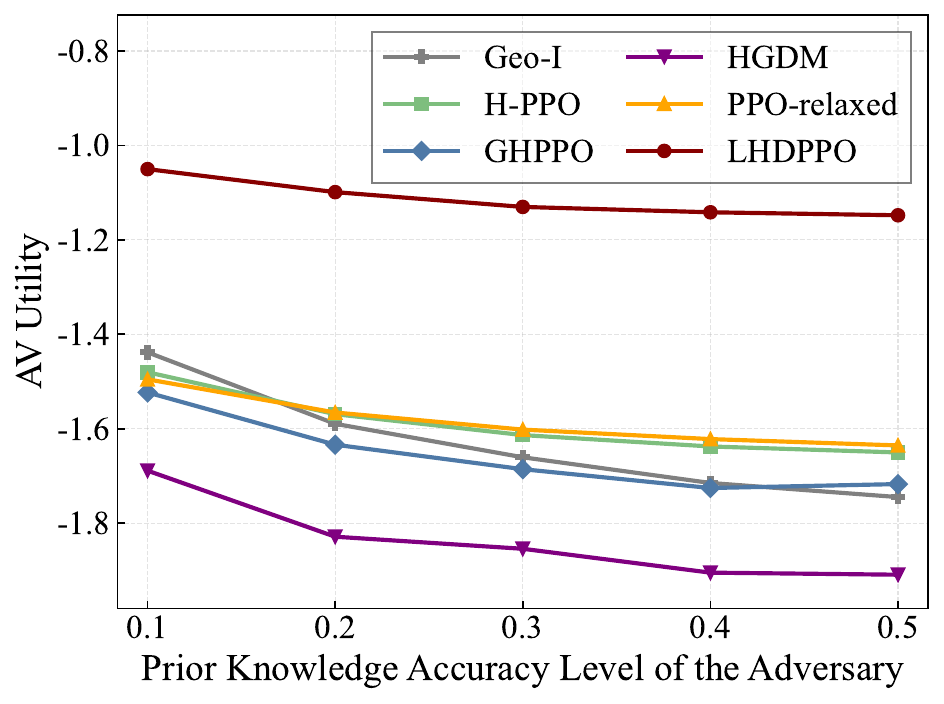}}
    \hfill
    \subfloat[\label{fig6:b}]{
    \includegraphics[width=0.32\textwidth]{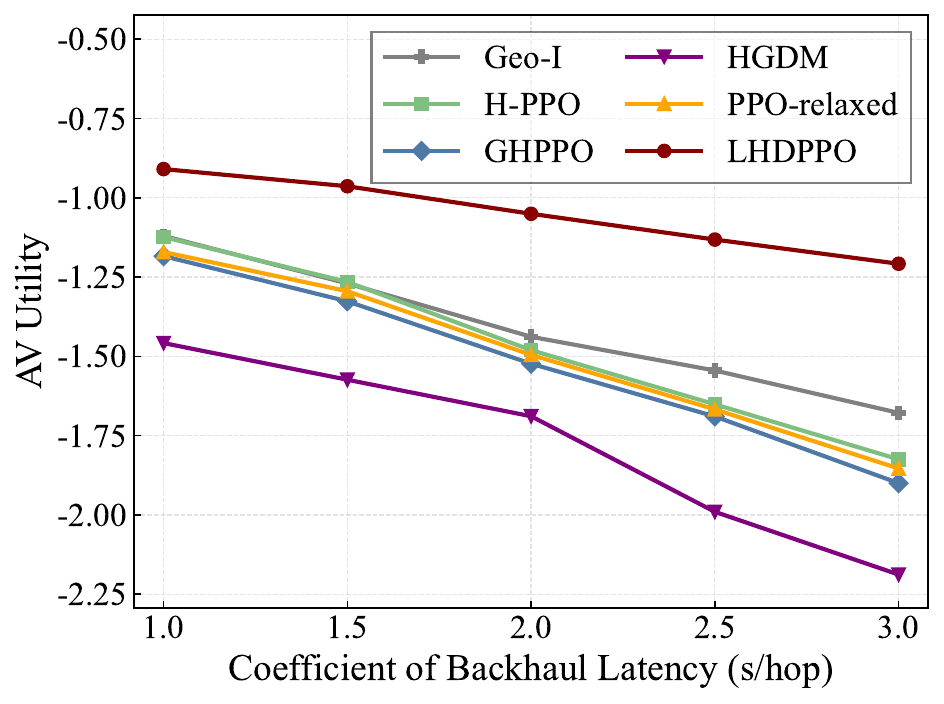}}
    \hfill
    \subfloat[\label{fig6:c}]{
    \includegraphics[width=0.32\textwidth]{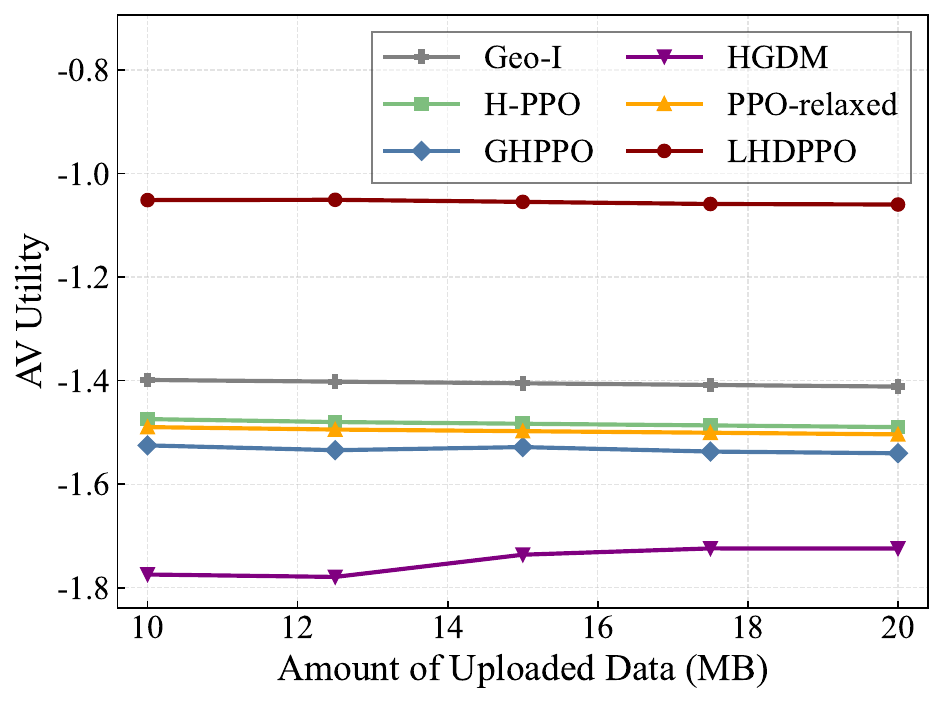}}
    \caption{(Color online) Sensitivity comparison of AV utility under different environment parameter settings. (a) Impact of adversary knowledge accuracy; (b) Impact of backhaul latency coefficient; (c) Impact of uploaded data size.}
    \label{fig6}
\end{figure*}

\textbf{Sensitivity analysis.} Figure~\ref{fig6} evaluates the sensitivity of AV utility with respect to three varying environment parameters, including adversary knowledge accuracy, backhaul latency coefficient, and uploaded data size. As shown in Figure~\ref{fig6:a}, increasing the adversary's prior knowledge accuracy strengthens the inference attack and reduces the utility of all schemes, but LHDPPO still achieves 29.8\% higher AV utility than the best baseline under the most challenging prior knowledge accuracy setting (i.e., $0.5$), showing that its hybrid policy can adaptively strengthen privacy protection when the adversary becomes more informed. Figure~\ref{fig6:b} shows that a larger backhaul latency coefficient increases the cost of cross-server AI agent migration, whereas LHDPPO still improves AV utility by 28.0\% when the coefficient reaches 3 by avoiding unnecessary migrations while preserving privacy-enhancing virtual-space decisions. Figure~\ref{fig6:c} further shows that when the uploaded data size increases to 20 MB, LHDPPO maintains a 24.9\% AV utility improvement, indicating that the proposed algorithm remains effective under heavier data loads and greater service-latency pressure.

\begin{figure*}[t]
    \centerline{\includegraphics[width=0.75\textwidth]{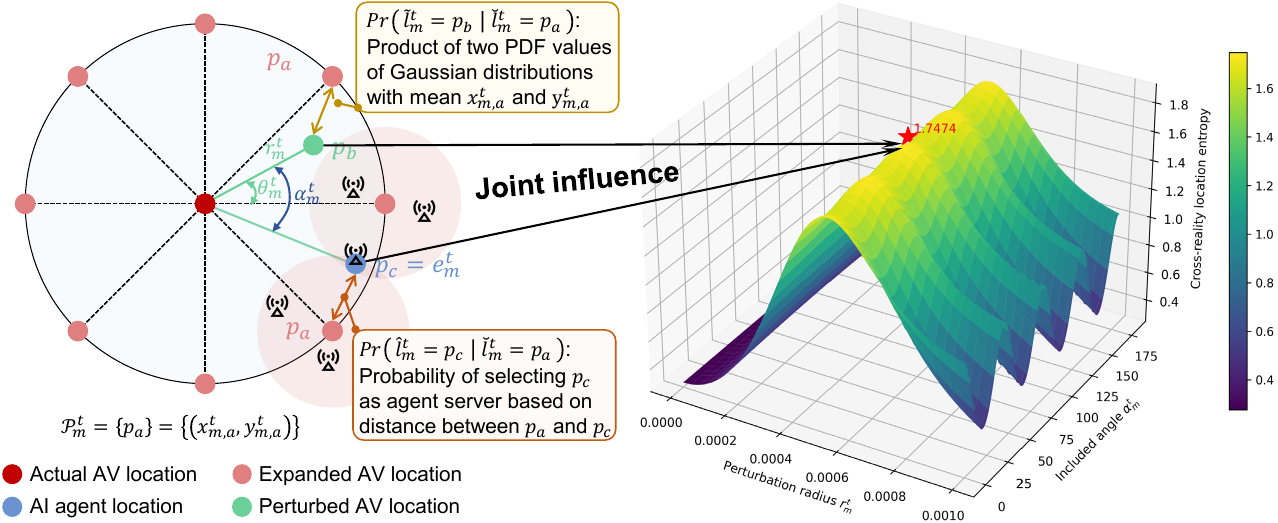}}
    \caption{(Color online) Example of the joint influence of hybrid action mechanisms for cross-reality location privacy protection.}
    \label{fig_LLM_influence}
\end{figure*}

\textbf{Ablation and LLM influence analysis.} To further investigate the effect of LLM-driven reward design, Table~\ref{tab:ablation} compares LHDPPO with HDPPO, which retains the dual-chain diffusion policy but removes the LLM-generated reward refinement module. Compared with HDPPO, LHDPPO reduces convergence steps by 14.6\%, improves AV utility by 21.5\%, improves cross-reality location entropy by 51.0\%, and reduces service response latency by 9.3\%, confirming that the LLM-generated reward provides useful guidance beyond diffusion-based hybrid action exploration. Figure~\ref{fig_LLM_influence} further illustrates how cross-reality location entropy varies with the hybrid action mechanism: the perturbation radius dominates the basic entropy trend, while the included angle $\alpha_m^t$, jointly determined by the perturbation direction and the selected agent server, provides additional privacy gains by changing the relative geometry between the released physical location and the virtual AI agent location. Therefore, by identifying this latent angle-related factor and incorporating it into reward shaping, the LLM-driven loop guides LHDPPO toward high-entropy hybrid actions rather than simply enlarging the perturbation distance.

\begin{table*}[t]
\centering
\caption{Ablation comparison between LHDPPO and HDPPO.}
\small
\label{tab:ablation}
\begin{tabular}{lccccc}
\hline
Scheme & Convergence Steps & AV Utility & Cross-reality Location Entropy & Service Response Latency & QoS Loss \\
\hline
LHDPPO & $4.1\times 10^6$ & $-1.073$ & $0.740$ & $115.995$ & $3.267$ \\
HDPPO & $4.8\times 10^6$ & $-1.367$ & $0.490$ & $127.851$ & $2.890$ \\
\hline
\end{tabular}
\end{table*}

\begin{figure}[t]
\centering
    \begin{minipage}[t]{0.28\textwidth}
      \centering
      \includegraphics[width=\linewidth]{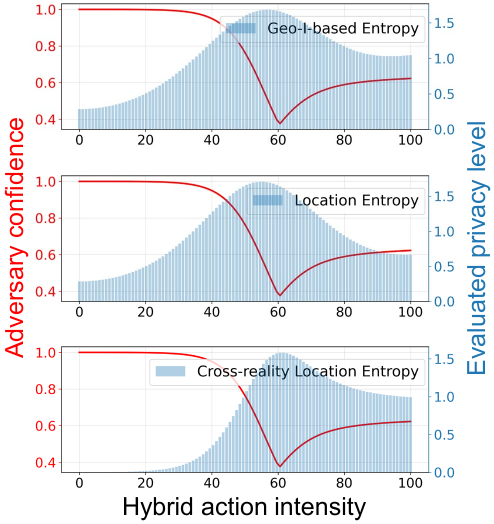}
      \captionof{figure}{(Color online) Comparison of the credibility of location privacy metrics under varying action intensities.}
      \label{fig7}
    \end{minipage}
    \begin{minipage}[t]{0.5\textwidth}
      \centering
      \includegraphics[width=\linewidth]{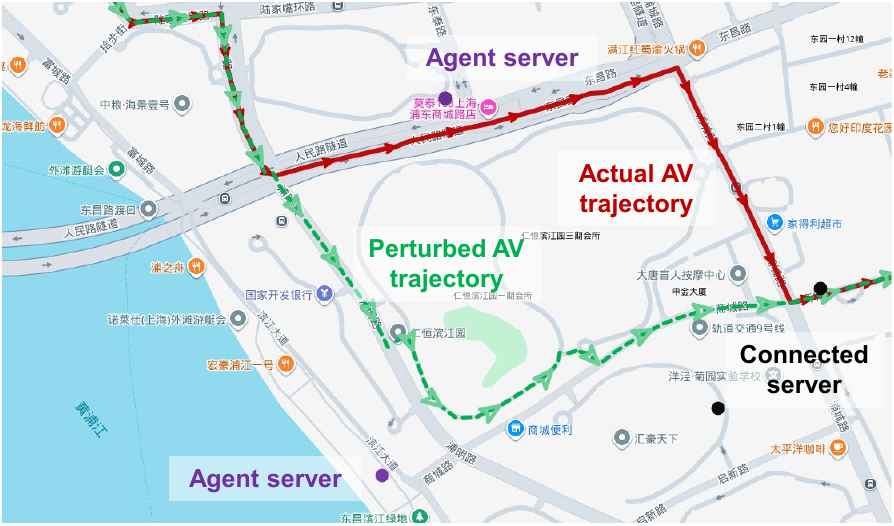}
      \captionof{figure}{(Color online) Application of the proposed framework in a real-world scenario. The figure visualizes the actual and perturbed trajectories of a target AV together with the locations of its agent servers.}
      \label{fig8}
    \end{minipage}\hfill
\end{figure}

\textbf{Privacy metric analysis.}
Figure~\ref{fig7} compares the evaluated privacy levels with the adversary confidence as the hybrid action intensity increases. We consider two representative baseline privacy metrics: Geo-I-based entropy for physical location perturbation~\cite{andres2013geo} and location entropy for virtual service migration~\cite{wang2023location}. A credible privacy metric should exhibit a clear inverse relationship with adversary confidence, but Geo-I-based entropy and migration-based location entropy fail to satisfy this requirement because their evaluated privacy levels are not consistently low when the adversary confidence is high. This is because both metrics capture only a single dimension of location uncertainty and fail to account for the cross-reality privacy risks jointly induced by AV location disclosure in the physical space and AI agent location exposure in the virtual space. In contrast, the proposed cross-reality location entropy explicitly models the coupling between physical and virtual observations, yielding privacy evaluations that remain aligned with adversary confidence and thus providing a more reliable reward signal for reinforcement learning.

\textbf{Case study in real scenarios.}
Figure~\ref{fig8} presents a real-map case study to illustrate how the proposed framework protects cross-reality location privacy in practice, where the red solid line denotes the actual AV trajectory and the green dotted line denotes the perturbed trajectory generated by the learned policy. It can be observed that the learned policy produces reasonable perturbations along the route while dynamically selecting agent servers around the neighborhoods of both the actual and perturbed trajectories, rather than fixing the AI agent on a certain nearby server. This coupled physical-virtual strategy prevents a stable cross-reality alignment between the reported physical trajectory and the observed AI agent deployment, thereby reducing adversary confidence while maintaining practical service quality in real scenarios.

\section{Conclusion}
\label{Conclusion}
This paper has proposed a cross-reality location privacy protection framework for agentic AI-driven AVs in 6G-enabled vehicular metaverses. The framework jointly optimizes continuous location perturbation and discrete privacy-aware AI agent migration, in which a cross-reality location entropy metric is designed to quantify AV position uncertainty under hybrid actions. Combining location privacy, service latency, and QoS loss, a non-convex mixed-integer programming problem has been established for optimizing the hybrid action mechanism. To address the formulated problem, we have developed an LLM-enhanced Hybrid Diffusion Proximal Policy Optimization (LHDPPO) algorithm, which combines LLM-driven informative reward design with dual-chain GDM-based policy exploration and PPO-based stable policy updates to improve learning performance and scalability. Extensive experiments on real-world mobility and base station datasets have demonstrated that the proposed framework effectively protects location privacy while maintaining user immersion, indicating its practical applicability for large-scale 6G-enabled vehicular metaverse deployments. Since correlated mobility patterns may lead to lower cross-reality location entropy than random mobility patterns, future work will investigate adaptive privacy protection under heterogeneous AV mobility patterns and stronger adversaries with heterogeneous background knowledge.

\Acknowledgements{This work was supported by the National Natural Science Foundation of China (Grant Nos. 62572132 and U25A20464), Guangdong Basic and Applied Basic Research Foundation (Grant No. 2025A1515010137), and the National Research Foundation of Korea (NRF) grant funded by the Korea government (MSIT) (Grant No. RS-2026-25470140).}

\end{document}